\documentclass[11pt]{article}
\usepackage{latexsym}
\usepackage{amsmath}
\usepackage{amssymb}
\usepackage{epsfig}
\usepackage{url}
\usepackage{xspace}

\usepackage{psfrag}

\usepackage{rotating}

\usepackage{ulem}

\usepackage{soul}

\RequirePackage{natbib}
\newcommand{\myprime}{\,\!'}
\newcommand{\myt}{\,\!^{(t)}}

\newcommand{\iaug}{i^*}
\newcommand{\ivec}{{i^\prime}}

\newcommand{\paua}{pa(u)}
\newcommand{\pava}{pa(v)}

\newcommand{\pa}{pa}

\newcommand{\pav}{pa(v)}

\newcommand{\be}{\begin{equation}}
\newcommand{\ee}{\end{equation}}
\newcommand{\ben}{\begin{eqnarray}}
\newcommand{\een}{\end{eqnarray}}

\newcommand\ind{\bot\hspace*{-6pt}\bot}
\newcommand\jind[2]{#1\ind#2}

\newcommand {\vet}{\mbox{vec}}

\newcommand {\dn}[1] {\mbox {\boldmath$#1$} }

\newcommand {\etal} {{\it et al.}\xspace}

\usepackage{color}

\usepackage{graphicx}

\usepackage{soul}

\title{Probability Based Independence Sampler  for \\Bayesian Quantitative Learning
in \\ Graphical Log-Linear  Marginal Models}

\author{Ioannis Ntzoufras
\\
{\it \small Department of Statistics, Athens University of Economics and Business, Greece}
\and Claudia Tarantola
\thanks{Address for correspondence: Claudia Tarantola, Department of Economics and Management, University of Pavia, Pavia, Italy.\texttt{E-mail:~claudia.tarantola@unipv.it}
}\\
{\it \small Department of Economics and Management, University of Pavia, Italy}
\and Monia Lupparelli\\
{\it \small Department of Statistical Sciences, University of Bologna, Italy}}

\begin{document}

\maketitle

\begin{abstract}

 Bayesian methods for graphical log-linear marginal models   have not
been developed in the same extent as traditional frequentist approaches.
In this work, we introduce a novel Bayesian approach
for quantitative learning for such models. These models belong to curved
exponential families that are difficult to handle from a Bayesian perspective.
Furthermore, the likelihood cannot be analytically
expressed as a function of the marginal log-linear interactions,
but only in terms of cell counts or probabilities.
 Posterior
distributions cannot be directly obtained, and MCMC methods are
needed. Finally, a well-defined model requires parameter values that lead to
compatible marginal probabilities. Hence, any MCMC should account for this important restriction.
We construct a fully automatic and
efficient MCMC strategy for quantitative learning for graphical
log-linear marginal models that handles these problems.
While the prior is expressed in terms of the marginal log-linear interactions, we build an MCMC algorithm that employs a proposal on
the probability parameter space. The corresponding proposal on the
marginal log-linear interactions is obtained via parameter
transformation.
 By this strategy, we achieve to
move within the desired target space. At each step, we directly work
with well-defined probability distributions.
 Moreover, we can exploit
a conditional conjugate setup to build an efficient proposal on probability parameters. The proposed
methodology is illustrated by a simulation study and a real dataset.

\end{abstract}

\par



\section{Introduction}

Statistical models which impose restrictions on marginal distributions of categorical data have received considerable attention especially in social and
 economic sciences; see, for example, in \cite{Bergsma_et_al}.
 A particular appealing class is that of  log-linear marginal models introduced by \cite{BergsmaRudas}, that includes as special cases log-linear and multivariate logistic models.
The marginal log-linear  interactions are estimated using the frequencies
of appropriate marginal contingency table, and    expressed in terms of log-odds ratios.
This setup is important in cases where information is available for specific marginal associations via odds ratios  (i.e. marginal log-linear interactions) or when partial information (i.e. marginals) is available.

 Log-linear marginal models have   been used to provide parameterisations for discrete graphical models; see
    \cite{Lupparelli_et_al},  \cite{Rudas_et_al} and \cite{EvansRichardson}.
    In particular, \cite{Lupparelli_et_al} used them
     to  define a parameterisation for discrete graphical models of marginal independence represented by a bi-directed graph.
The absence of an edge in the bi-directed graph indicates marginal independence, and  the corresponding marginal log-linear interactions (i.e. the corresponding log-odds ratio) are constrained to zero.

Despite the increasing interest in the literature for graphical
log-linear marginal models, Bayesian analysis has not been
developed as much as traditional methods. Some context specific
results have been presented by e.g. \cite{SilvaGhahramania},
\cite{Bartolucci_et_al} and  \cite{NtzoufrasTarantola2013}.
For graphical log-linear marginal models, no conjugate analysis is available.
Therefore, Markov chain Monte Carlo (MCMC) methods must be employed.
Nevertheless, the likelihood of the model cannot be analytically expressed as a function of the marginal log-linear interactions.
This creates additional difficulties on the implementation of MCMC methods since, at each step,  an iterative procedure needs to be applied in order to calculate the cell probabilities and consequently the model likelihood.
Moreover, in order to have a well-defined model of marginal independence,
we need to construct an algorithm which generates parameter values that lead to a joint probability distribution with compatible marginals.
To achieve this, we need an MCMC scheme which moves within the restricted space of parametrisations satisfying the conditions induced by the compatibility of marginal distributions.

In this paper we construct a novel, fully automatic, efficient MCMC strategy for quantitative
learning for graphical log-linear marginal models that handles the previously discussed problems.
We assign a suitable prior distribution on the marginal log-linear parameter vector, while
the  proposal is expressed in terms of the probability parameters.
The proposal distribution of marginal log-linear interactions is constructed by simply transforming
generated candidate values of probability parameters. The corresponding proposal density is directly available by implementing standard theory about functions of random variables.
The advantages of this strategy are clear: the joint distribution
factorises under certain conditional independence models, and the
likelihood can be directly expressed in terms of probability
parameters.
Furthermore, efficient proposal distributions can be
constructed applying the conditional conjugate approach of
\cite{NtzoufrasTarantola2013}, that exploit the representation of
the model in terms of   an augmented  Direct Acyclic Graph (DAG).
 We present two probability based samplers:
the probability-based independence sampler (PBIS) and
the prior adjustment algorithm (PAA).
The first one is an augmented Metropolis-Hasting algorithm, while the latter it is only an approximation of an independence Metropolis-Hastings algorithm. Hence, related theoretical results cannot be invoked directly  for it. Nevertheless, empirical comparisons between the two methods indicate that PAA provides results similar to the ones obtained via PBIS in a faster and in more efficient way.

Assigning  a  prior distribution on the marginal log-linear interactions
rather than on the probability parameters represents a novel approach
and is particularly handy in the presence of informative prior about odds for specific marginal associations.
For instance, symmetry constraints, vanishing
high-order associations or further prior information about the
joint and marginal distributions can be easily specified by
setting linear constraints on marginal log-linear terms instead of
non-linear multiplicative constraints on the probability space.
Finally, marginal log-linear parameters are  more appealing than probabilities because they are interpretable measures of association between observed variables.

The plan of the paper is as follows. In Section \ref{Sec_model}, we introduce
discrete graphical models of marginal independence and  the
 marginal log-linear parameterisation. In Section \ref{Bayesian_msu}, we describe
the considered prior set-up. Section \ref{our_mcmc} is devoted to the proposed MCMC
strategies.
 The methodology is
illustrated in Section 5 which presents  a simulation study and a real data analysis. In Section \ref{disc}, we conclude
with a brief discussion and ideas for future research.

\section{Model Specification and Parameterisation \label{Sec_model}}

In this section we briefly introduce discrete graphical models of marginal
independence, the related notation and terminology, and the
corresponding  marginal log-linear parameterisation.

\begin{figure}[ht]
\caption{Example of  bi-directed graph}
\label{bidir}
  \centering  \includegraphics[width=3cm]{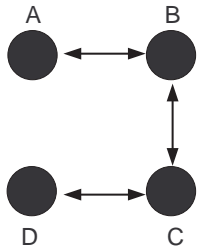}
          \end{figure}

 A bi-directed  graph $G=\left(\mathcal{V}, E\right)$, is a graph  with  vertex set ${\mathcal{ V}}$,
and  edge set $E$, such that  $(u,v) \in E$ if and
only if  $(v,u) \in E$. Following   \cite{Richardson} edges are represented via  bi-directed arrows.
An alternative representation, proposed by  \cite{CoxWermuth}, is by undirected
dashed edges.
The skeleton $\overline{G}$ of a bi-directed graph
$G$ is the graph obtained by making all edges undirected. A $\vee$ configuration
is every triplet of vertices $(u,v,z)$
in  $\overline{G}$ with edges $(u,v)$ and $(v,z)$
and with no edge connecting $u$ and $z$.
A vertex set is connected if there is a path between every pair of  vertices belonging to it.
We consider a set of random variables
$Y_{\cal V}= \big( Y_v,~ v\in {\cal V} \big)$,
each one taking values $i_v \in {\cal I}_{v}$; where ${\cal I}_{v}$ is the set of possible levels for
variable $v$. The cross-tabulation of variables  $Y_{\cal V}$
produces a $|{\cal V}|$-way contingency table with cell
frequencies ${\dn n} = \big( n(i), ~ i \in {\cal I} \big)$ where
${\cal I}={\mbox{\Large \dn{\times}}}_{v\in {\cal V}} \ {\cal I}_v$.
We  further assume that
$ \dn{n}  \sim Multinomial\big( \dn{p}, N  \big) $ with
$\dn{p}=\big( p(i), ~ i \in {\cal I} \big)$;
$p(i)$ is the joint probability for cell $i \in {\cal I}$, and $N=\sum \limits_{i \in {\cal I}} \ n(i)$.
A bi-directed graph $G$ is  used to represent marginal
independencies between variables  $Y_{\cal V}$ which are expressed
as non-linear constraints over the set of the joint probabilities
$\dn{p}$. The list of independencies implied by a bi-directed
graph can be obtained using the pairwise Markov property  \citep{CoxWermuth} and the connected set Markov property
\citep{Richardson}. For discrete variables the connected set
Markov property implies the pairwise Markov property, whereas the
converse is not generally true. Following \cite{DrtonRichardson}, we define a discrete graphical model of marginal independence as the family of probability distributions for
$Y_{\cal V}$ that satisfy the connected set Markov  property.
For example the bi-directed graph in  Figure \ref{bidir} encodes the marginal
independencies  
$\jind{Y_{\{A,B\}}}{Y_{D}}$ and  $\jind{Y_{A}}{Y_{\{D,C\}}}$ under the connect set Markov property.

The marginal log-linear parameterisation for bi-directed graphs
has been proposed by \cite{Lupparelli} and  \cite{Lupparelli_et_al}; it is  based on the class of  log-linear marginal models
of \cite{BergsmaRudas}.
According to \cite{BergsmaRudas} the parameter vector $\dn{\lambda}$, containing the marginal log-linear interactions, can be obtained as
  \be
 \dn{\lambda} = \dn{C} \log \big( \dn{M} \dn{P} \big) \mbox{~with~} \dn{P} = \vet(\dn{p})
 \label{lambda_of_marginal_model},
\ee where $\mbox{vec}(\dn{p})$ is a vector of dimension $|{\cal
I}|$ obtained by rearranging the elements $\dn{p}$ in a reverse
lexicographical ordering of the corresponding variable levels, with
the level of the first variable changing first.  Each marginal
log-linear interaction  satisfies identifiability constraints (here sum-to-zero constraints), and this is achieved via an
appropriate contrast matrix $\dn{C}$.  The interactions are
calculated from a specific marginal table identified via the
marginalisation matrix
 $\dn{M}$.
 They are characterised  by two sets of variables: one
set that refers to the marginal table in use and a second set (subset of the first one) that identifies which variables are involved
 in the interaction of interest.
  Finally, the first order interactions correspond to the main effects.
   We denote with  $\dn{\lambda}^{M_m}$  the
parameter vector containing all interactions estimated
 from the marginal probability table $\dn{p}^{M_m}$.
 Matrices $\dn{C}$ and $\dn{M}$ define a smooth mapping between the probability space and the marginal log-linear space;
see \cite{forcina2010} for technical details about the smoothness.
Matrix $\dn{C}$   controls the induced log-linear parameterisation.
In this work we have used the sum-to-zero parameterisation which is one of the possible configurations leading to different sets of log-linear parameters.
Details  for the  the construction of  $\dn{C}$ and $\dn{M}$ are
available in Appendices \ref{calculation_M},
  \ref{calculation_C}  and in Section \ref{secprior}.

A graphical model of marginal independence is defined by zero
constraints on specific marginal log-linear interactions. More
precisely, we apply the following procedure presented by
\cite{Lupparelli} and  \cite{Lupparelli_et_al}: (i) define a
hierarchical ordering (see \citealp{BergsmaRudas})
of the marginals  corresponding to  disconnected sets of the bi-directed
graph; (ii) append the marginal corresponding to the full table at
the end of the list if it is not already included; (iii) for every
marginal table estimate all  interactions that have not been
already obtained  from the marginals preceding it in the ordering;
(iv) for every marginal table corresponding to a disconnected set
of $G$, restrict the highest order log-linear interaction
 to zero.

The graphical structure imposes constraints of the type
$$
\dn{K} \log \Big( \dn{M} \dn{P} \Big)  = 0
$$
with $\dn{K}$ being the sub-matrix of $\dn{C}$ for which the corresponding elements of
$\dn{\lambda}$ are restricted to zero.
This parameterisation depends on the ordering of the marginals selected in step (i).
 Furthermore, it does not always satisfy variation independence; see \cite{Lupparelli_et_al},
\cite{Rudas_et_al}, and \cite{EvansRichardson}.
If the marginal selected in step (i) are order decomposable then variation independence is guaranteed \citep{BergsmaRudas} so that the  marginal-log linear parameter space is rectangular.
Hence, in this paper we focus  on models based on an order decomposable set of marginals.
Graphical log-linear marginal models based on  bi-directed graphs with three and four vertices are   always variation independent whichever the
chosen ordering of the marginals.

\section{Bayesian Model Set-up \label{Bayesian_msu}}

\subsection{Prior Specification for Marginal Log-linear Interactions \label{secprior}}

 In order to specify our prior setting for  graphical marginal log-linear models  it is useful to rewrite the model described by equation
  (\ref{lambda_of_marginal_model}) in the following extended form
\be
\left(  \begin{array}{l} \dn{\lambda}^{M_1} \\ \vdots \\ \dn{\lambda}^{M_m}  \\ \vdots \\ \dn{\lambda}^{|\cal{{ M}}|} \end{array}  \right)
       =\mbox{diag}\left( \dn{C}_1, \cdots, \dn{C}_m, \cdots, \dn{C}_{|{\cal M}|}  \right)
        \left(  \begin{array}{l} \log\dn{P}^{M_1} \\ \vdots \\  \log\dn{P}^{M_m}  \\ \vdots \\  \log\dn{P}^{|\cal{{ M}}|} \end{array}  \right),
        \label{expanded_model}
\ee where  ${\cal M}=\{ M_1,  \dots, {M_m}  \dots, M_{|{\cal M}|} \}$ is the set of marginals under consideration, $\dn{\lambda}^{M_m}$ is the
parameter vector
obtained from the marginal probability table $\dn{p}^{M_m}$
which is re-arranged to a vector denoted by $\dn{P}^{M_m}$ for all $m=1,2,\dots, |{\cal M}|$.
  The marginals under consideration correspond to the disconnected sets of the graph. If the graph is connected,
we need to append   the margin that corresponds to the full table to the previous set of marginals.
The contrast matrix $\dn{C}$ is  a block diagonal matrix with elements $\dn{C}_m$.
Each sub-matrix $\dn{C}_m$ is obtained  by inverting the design matrix
${\mathbf X}_{M_m}$ of the saturated model fitted on
marginal $M_m$, and deleting rows corresponding to
interactions that are not estimated from that specific marginal
table; see  Appendix \ref{calculation_C} for details. 
From (\ref{expanded_model}), we directly obtain that the interactions $\dn{\lambda}^{M_m}$ of the marginal $M_m$ are obtained by
$$
\dn{\lambda}^{M_m} = \dn{C}_m \log\dn{P}^{M_m}
\mbox{~for all~} M_m \in {\cal M}.
$$
Every $\dn{\lambda}^{M_m}$ may contain
interactions that are constrained to zero due the graphical
structure $G$ and the induced contrast matrix.
In the following we
focus only on non-zero elements of  ${\dn{\lambda}}$,  on which we assign a suitable prior distribution. We denote by
$\vec{\dn{\lambda}}$ the
set of elements of ${\dn{\lambda}}$  not
restricted to zero, that is
\begin{eqnarray*}
\vec{\dn{\lambda}} &=&  \left( \vec{\dn{\lambda}}^{M_m}; M_m \in {\cal M}  \right)
\mbox{~with~}
\vec{\dn{\lambda}}^{M_m} = \left( \lambda^{M_m}_\jmath :  \lambda^{M_m}_\jmath \neq 0, \jmath=1,\dots,  \textsc{r}_{C_m} \right),
\end{eqnarray*}
where $\textsc{r}_{C_m}$ is the number of rows of the contrast matrix $\dn{C}_m$ for marginal $M_m \in {\cal M} $.

When no information is available about $\vec{\dn{\lambda}}$,
we can work separately on each element of $\vec{\dn{\lambda}}$, assigning suitable  independent normal
prior distributions
 with large variance to express ignorance, i.e.
$$
f(\vec{\lambda}_\jmath) \sim N( 0, \sigma_{\jmath}^2 ) \mbox{ for }
\jmath=1,2,\dots, d_{\vec{\lambda}},
$$
where $d_{\vec{\lambda}}$ is the  number of  elements of $\vec{\dn{\lambda}}$.

A more sophisticated approach can be  based on the prior suggestion
of \cite{DellaportasForster}  for standard log-linear models
of the form $\log \dn{\mu} = \dn{X}  \dn{\beta}$.
They  considered the following prior distribution
\begin{equation}
\dn{\beta} \sim N \Big( \dn{\theta}, \, 2 |{\cal I}| \big( \dn{X}^T \dn{X} \big)^{-1} \Big)
\mbox{~with~} \dn{\theta} = ( \log \overline{\dn{n}}, \, 0, \dots, \, 0 )^T,
\label{DF_prior}
\end{equation}
where $\dn{\mu}$ is the  vector of the expected number of cell frequencies and
$|{\cal I}|= \prod \limits_{v \in {\cal V}} \ |{\cal I}_v|$ is the number of cells of the contingency table.
Matrix  $\dn{X}$ represents the design matrix of the model; see Appendix \ref{calculation_C} for more details and full specification.

This prior was suggested as a default choice for model comparison of log-linear models
and it arises naturally for log-linear marginal models since within each marginal we essentially
work by fitting standard log-linear models.
 In fact, following the procedure described in Section  \ref{Sec_model},
for every marginal we fit a saturated log-linear model, and then we keep only  interactions that have not been already estimated from the marginal preceding it in the examined ordering.

 Dellaportas and Foster prior was obtained after matching the prior moments  of other default priors
 used for the probability parameters in conjugate analysis of graphical models such as the Jeffreys and the Perks prior distributions, for more details  see  Sections 3.2 and 3.3 in \cite{DellaportasForster}.
 Moreover,  the prior distribution induced for the log-odds ratios remains the same even if extra, marginally independent, categorical variables are added in the model.
Finally, such prior arises naturally in our context since the adopted parameterisation  allows us to work by fitting standard log-linear models within each marginal.
These are the main reasons why we recommend  adopting this prior here.

The prior mean vector $\dn{\theta}$
has all its elements equal to zero except the first one which is equal to the logarithm of the average number of observations per cell $\overline{\dn{n}}$.
Under sum-to-zero constraints, $  \dn{X}^T \dn{X}  $ is a block diagonal matrix resulting to
a set of independent priors for interactions referring to different set of variables.

In order to construct the prior distribution on $\vec{\dn{\lambda}}$,
we   work separately  on each single set ${\dn{\lambda}}^{M_m}$
obtained from marginal $M_m$ and we proceed as follows.
Let $\dn{\lambda}^{M_m}_S$ be the parameter vector for the saturated model that can be estimated from marginal $M_m$; by construction, it  coincides with the parameter vector  of the
saturated standard log-linear model obtained from this marginal.
In terms of \cite{DellaportasForster} parameterisation,
$\dn{\lambda}^{M_m}_S$ can be written as
$\dn{\lambda}^{M_m}_S = \dn{\beta}^{M_m} - \log(N) \, {\mathbf X}_{M_m}^{-1} \dn{1}$.
Hence, from (\ref{DF_prior}), the default prior for
$\dn{\lambda}^{M_m}_S$ is given by
\be
\dn{\lambda}^{M_m}_S
\sim  N\Big( \dn{\theta} - \log(N) \, {\mathbf X}_{M_m}^{-1} \dn{1}\;,\;
              2 |{\cal I}_{M_m}| \left( \dn{X}_{M_m}^T \dn{X}_{M_m} \right)^{-1}\Big).
\label{parameterprior}
\ee
The prior of $\vec{\dn{\lambda}}^{M_m}$ is obtained via marginalisation from (\ref{parameterprior})
since $\vec{\dn{\lambda}}^{M_m}$ is a subset of $\dn{\lambda}^{M_m}_S$.
When using  sum-to-zero constraints,
then $\dn{X}_{M_m}^{-1}\dn{1}= (1,0,\dots,0)^T$
resulting to a prior mean equal to $\theta_\jmath=0$ for all  $\vec{\dn{\lambda}}^{M_m}_\jmath$
except for the intercept for which the prior mean is given by $\log \overline{\dn{n}} -\log(N)$.
This prior is greatly simplified to a product of independent $N(0,2)$ for all marginal log-linear interactions
(except for the intercept) in the case of binary variables. 
Finally, the prior for the full parameter vector $\vec{\dn{\lambda}}$ is
obtained as a product of the priors on $\vec{\dn{\lambda}}^{M_m} $.

\subsection{Likelihood Specification and Posterior Inference}

The likelihood cannot directly be expressed in terms of $\dn{\lambda}$
(or equivalently $\vec{\dn{\lambda}}$) but only as a function of the probability parameter
\begin{eqnarray}
 f(\dn{n}|\dn{\lambda}) &=&
\frac{\Gamma(N+1)} {\prod \limits _{i \in {\cal I}}
\Gamma\big(n(i)+1\big)}
     \prod_{ i \in {\cal I}}   \wp_i(\dn{\lambda})  ^{n(i)},   \nonumber \label{loglike_lambda} \\
\mbox{where\;} \wp_i(\dn{\lambda}) & \equiv & \left\{ p(i) :  \dn{\lambda} = \dn{C} \log \Big( \dn{M} \dn{P}  \Big) \right\}, \mbox{~for all~} i \in {\cal I}~.
\label{like_actual_1}
\end{eqnarray}
Unfortunately, in order to obtain $\wp_i(\dn{\lambda})$, or equivalently $\dn{P}$, from  (\ref{lambda_of_marginal_model}),
 we need to implement an iterative algorithm,
and therefore the likelihood cannot be written in a closed form
expression, see, for example, \cite{BergsmaRudas}, and \cite{Lupparelli_et_al}. As a consequence of this,  the corresponding posterior
distribution of $\lambda$ (or equivalently
 $\vec{\dn{\lambda}}$), cannot be  evaluated straight away.
Hence, MCMC methods are needed for posterior inference of $\vec{\dn{\lambda}}$.

   Metropolis-Hastings scheme on the marginal log-linear interactions   can be implemented to estimate the posterior distributions of the  parameter vector. Nevertheless,
such an algorithmic strategy will be inefficient since in
 every MCMC iteration, the joint probabilities  corresponding to the proposed values of
$\dn{\lambda}$ need to be calculated using an iterative algorithm.
This will considerably slow down the MCMC sampler.
The use of the iterative procedure  may also result to unstable solutions for specific combinations of lambda values. During simulations we  have encountered several inconsistencies in the
estimates for specific configurations  that may have unpredicted effect on the MCMC runs and
the corresponding estimated posterior distributions.
Another important issue is the fact that
the constructed  algorithm should generate parameter values that lead to well-defined joint probability distributions with compatible marginals.
Defining a  proposal distribution
that allows  moving within the space of
variation independent log-linear marginal models is still an open issue.

For the above reasons, in the following section, we propose a novel
 MCMC strategy based on the probability representation
of the model.  Such probabilities are compatible by construction, the corresponding marginal log-linear interactions will be also compatible.
 For comparative purposes, we have implemented a ``vanilla'' random walk algorithm MCMC  on $\vec{\dn{\lambda}}$ (referred to as \mbox{RW-$\lambda$}),
which proposes to change each log-linear parameter vector independently for every single marginal based on a decomposable ordering.



\section{Probability Based MCMC Samplers}
\label{our_mcmc}

\subsection{Initial Set-up and Data Augmentation for MCMC}
\label{sec41}

\begin{figure}[hbtp]
      \caption{Bi-directed  4-chain graph and the corresponding Markov equivalent DAG over the observed margin. }
      \label{fig_chain}
      \begin{center}
      \begin{tabular}{c@{\hspace{3cm}}c}
      \includegraphics[width=3cm]{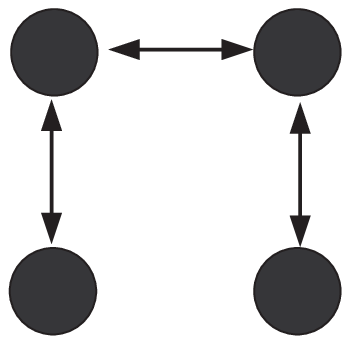} &
      \includegraphics[width=3cm]{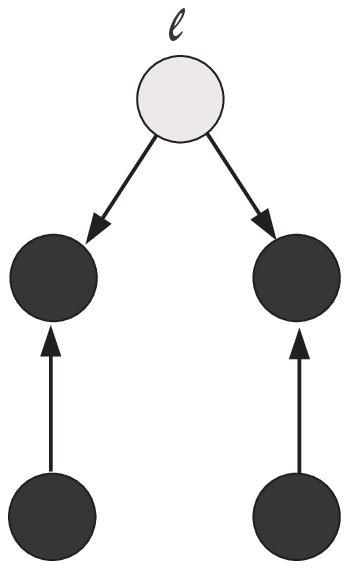} \\
      (a) Bi-directed graph & (b) DAG representation
      \end{tabular}
      \end{center}
   \end{figure}

 Following the notation of \cite{NtzoufrasTarantola2013}, we can
divide the class of graphical log-linear marginal
 models in two major categories: homogeneous and non-homogeneous models.
 A bi-directed graph is named homogeneous if it does not include a
bi-directed 4-chain or a chordless 4-cycle as sub-graphs.
 Both type of models are shown to be compatible, in terms of independencies, with a certain DAG
 representation (augmented DAG);
 see \cite{PearlWermuth}, \cite{{DrtonRichardson}} and \cite{SilvaGhahramania,SilvaGhahramanib}.
 Nevertheless, while homogeneous models  can be generated  via a  DAG with the same  vertex set, for   the non-homogeneous ones  it is necessary to include
   some additional latent variables  which  makes Bayesian inference challenging.
  The advantage of the augmented DAG representation is that the joint probability over the augmented variable space (including    both observed and latent variables) can be written using the standard  DAG  factorisation.
    An example of the  DAG representation with latent variables for a non-homogeneous bi-directed graph is provided in Figure \ref{fig_chain}.

In order to construct  the augmented DAG, with the minimal set of latent variables,   we apply the following  procedure presented in \cite{PearlWermuth}.
Given the skeleton $\overline{G}$ of the examined graph, we assign
arrows $v_\imath\longrightarrow v_\jmath \longleftarrow v_k$ to
each $\vee$ configuration $(v_\imath,v_\jmath,v_k)$ in
$\overline{G}$, constructing in this way the sink orientation of $G$.
If  no edge in the  sink orientation  is bi-directed we consider an acyclic
orientation of the undirected edges.
If the sink orientation contains bi-directed edges,  we  substitute every bi-directed edge $v_1\longleftrightarrow v_2$ with
the directed configuration
$v_1\longleftarrow \ell \longrightarrow v_2$, where vertex $\ell$
represents a hidden or latent variable. 
Finally, a DAG   is constructed via an acyclic
orientation of the undirected edges which are present in the sink orientation of the graph.
 Any  DAG which encodes the same independencies between the observable variables as  in $G$ is called augmented DAG of $G$.

   \cite{NtzoufrasTarantola2013} exploited the connection between bi-directed graphs and DAGs to
develop a Gibbs sampler based on a probability parameterisation of the model.
 They parameterised the augmented
 DAG model  in terms of a set of
marginal and conditional probability parameters
$\dn{\Pi}$, on which they
implemented a conjugate analysis based on products of Dirichlet
distributions.
For homogeneous bi-directed graphs, the posterior distribution of the  probability vector for  the observed variables can be obtained analytically.
 On the other hand, for non-homogenous graphs,
a Gibbs sampler can be obtained by calculating appropriate contrasts of the logarithms of the probability parameters
within the proposed data augmentation scheme.
 The Gibbs sampler for the non-homogenous models is essentially composed by three  steps:
\begin{description}
\item[Step 1:] Generate random splits for each bi-directed edge in the sink orientation of $G$ by using   a   multinomial distribution (i.e. we generate the latent data).

\item[Step 2:] Generate random samples from the induced Dirichlet posterior conditional distributions
for each set of probability parameters of the augmented DAG, given the augmented table of Step 1.

\item[Step 3:] Calculate the joint and the marginal probabilities for the observed variables
by implementing the appropriate marginalising transformations to the probability parameter values of Step 2.

\end{description}
Additionally, the posterior distribution of the log-linear interactions can be obtained as  by-product of this approach. This can be achieved by calculating appropriate contrasts of the logarithms of the joint probability parameters obtained in Step 3.
We may implement this by either adding an additional step to the above algorithm or by implementing the induced transformation on the final MCMC output of the algorithm.

 However this approach is open to the following critics.
First of all,
prior information is usually available about the marginal association of the observed variables.
This is typically expressed by the size of log-linear interactions which correspond to log-odds.
Any prior information available for log-odds cannot be incorporated in  a probability
based method in a trivial manner.
For instance, symmetry constraints, context specific independence assumptions, vanishing high-order associations or further prior information about the joint and marginal distributions  can be  specified by setting zero or linear constraints on the marginal log-linear parameter space, instead of non-linear multiplicative constraints on the probability space.
  Specifying sub-models based on a priori knowledge is an important issue for discrete bi-directed graph models. In fact, they generally require to extimate a higher number of parameters compared to models of conditional independencies, such  as undirected graph and DAG models; see \cite{Lupparelli_et_al} for a discussion which compares the number of parameters in discrete bi-directed and undirected graph models with  the same skeleton. From this perspective, the marginal log-linear parameterisation  is definitively more appealing than the probability based approach.

  In this paper, we introduce a novel MCMC strategy, in which the model and the prior are expressed in terms of the marginal log-linear interactions, while the proposal is defined  on the probability parameter space.
By this way, we can incorporate any source of prior information that might be available
regarding marginal associations by  specifying the prior values of selected log-odds ratios (i.e. interactions).
Moreover, we can exploit the advantages  of  the conditional conjugacy of the algorithm of \cite{NtzoufrasTarantola2013} in order to obtain efficient proposals based on probability parameters.
 This approach allows us to control the induced values and constraints on interactions in the log-linear
space in more natural way than working in the probability space induced by the (augmented) DAG.

  Our approach does not account for any additional non-independence constraints  which may
arise by marginalising the latent variables from the augmented DAGs.
  Defining the exact class of marginal DAGs still represents an open issue; see Evans (2016) who discusses possible approximations. Moreover, these inequality constraints are not of great interest as they do not have a straightforward interpretation.
  Finally, it is worth noting that if  non-independence constraints
  are included after marginalising over the latent variable, the resulting likelihood is a close approximation of the true
  one.

Using the methodology of \cite{NtzoufrasTarantola2013}, we construct an MCMC sampler based
on proposing values in terms of joint probabilities, avoiding compatibility problems.
If the model is homogeneous  the dimension of $\dn{\Pi}$ is the
same as the dimension of $\vec{\dn{\lambda}}$. This is not true
for   non-homogeneous models since the dimension of $\dn{\Pi}$ is
greater than the dimension of $\vec{\dn{\lambda}}$. In this case
we need to augment the parameter space in order to implement
Metropolis-Hastings algorithm.
In the following, we denote the augmented set of marginal
log-linear interactions by with $\dn{\lambda}^{\cal A}$.
If the model is homogeneous, then $\dn{\lambda}^{\cal A} =  \vec{\dn{\lambda}}$.
Otherwise, we set $\dn{\lambda}^{\cal A} = ( \vec{\dn{\lambda}}, \dn{\xi})$;
where $\dn{\xi}$ is selected to be the last $d_\xi$
elements of   $\dn{\lambda}^{\cal A}$ with $d_\xi=\dim(\dn{\Pi})-\dim( \vec{\dn{\lambda}})$.
The vector $\dn{\xi}$  can be thought as an auxiliary set of parameters which are used to retain the dimension balance in the Metropolis-Hastings algorithm.

More precisely, for any graphical log-linear marginal
 model  $G$
we can obtain a Markov equivalent DAG over the observed margins, denoted by $D_G$, with augmented vertex set ${\cal A} = \{ {\cal V}, {\cal L} \}$,
where ${\cal L}$ is the set of additional latent variables; if  $G$ is homogeneous then ${\cal L} = \emptyset$
and ${\cal A}={\cal V}$.
Under this approach, the joint probabilities can be written as
\begin{eqnarray}
&&p(i) =  \sum _{ i_{\cal L} \in {\cal I}_{\cal L} } p^{{\cal A}}( i, i_{\cal L} )   \mbox{~for~} i \in {\cal I}_{\cal V},
\label{joint_probs} \\
&&p^{{\cal A}}( i, i_{\cal L} ) =p^{{\cal A}}( \iaug ) =\prod_{v \in \mathcal{A}}\pi_{v|\pav}\left(\iaug_v| \iaug_{\pav} \right) \mbox{~for~}
\iaug=(i,i_{\cal L})  \in {\cal I}_{\cal A},
\label{augm_joint_probs} 
\end{eqnarray}
and the probability parameter set is given by
$$
\dn{\Pi} = \mbox{vec}\Big( {\pi}_{v|\pav}( \iaug_v | \iaug_{\pav} ); \, \iaug_v \in {\cal I}_v \setminus \Big\{ |{\cal I}_v| \Big\},
\iaug_{\pav} \in {\cal I}_{\pav},  v \in {\cal A}  \Big),
$$
%
where
$p^{{\cal A}}( \iaug )=P\big( Y_{\cal V} = \iaug_{\cal V}, Y_{\cal L} = \iaug_{\cal L} \big)$ is the joint probability for
the  observed variables  $Y_{\cal V}$ and the latent variables  $Y_{\cal L}$,
$\pav$ stands for the parents set of  $v$ in graph $D_G$,
$\pi_{v|\pav}\big(\iaug_v|\iaug_{\pav} \big) = P\big( Y_v = \iaug_v | Y_{\pav}=\iaug_{\pav} \big)$
is the conditional probability of
each variable $Y_v$ given variables $Y_{\pav}$ in the parent set $\pav$ of $v$.
By using the induced augmented likelihood representation,
we are able to construct a Gibbs sampler based on
the conditional conjugate Dirichlet prior distributions on $\dn{\Pi}$
\citep{NtzoufrasTarantola2013}.

\subsection{The general algorithm}
\label{prior_adj}

The posterior distribution of the augmented set of marginal log-linear interactions is given by
\begin{eqnarray*} f( {\dn{\lambda}}^{\cal A} | \dn{n} ) &\propto&
f\Big( \dn{n} | {\mathbf \wp}({\dn{\lambda}}) \Big) f(
\vec{\dn{\lambda}} )  f(\dn{\xi}),
\end{eqnarray*}
where
$f(\dn{\xi})$ is a pseudo prior used for the additional parameters.

We consider a Metropolis-Hastings algorithm which can be summarised by the following steps.
\begin{enumerate}
\item [~] \hspace{-2em} For $t=1,\dots, T$, repeat the following steps:
\item Propose a new vector $\dn{\Pi}'$ from $q(\dn{\Pi}'|\dn{\Pi}^{(t)})$;
            where $\dn{\Pi}^{(t)}$ are the values of $\dn{\Pi}$ at $t$ iteration.

\item From $\dn{\Pi}'$, calculate the proposed joint probabilities $\dn{p}'$
      (for the observed table)
      using equations (\ref{joint_probs}) and (\ref{augm_joint_probs}).

\item From  $\dn{p}'$, calculate ${\dn{\lambda}}\myprime$ using (\ref{lambda_of_marginal_model})
            and then obtain the corresponding non-zero elements $\vec{\dn{\lambda}}\myprime$.

\item Set $\dn{\xi}' = \dn{\Pi}_{\xi}'$; where $\dn{\Pi}_{\xi}'$ is a pre-specified subset of $\dn{\Pi}'$ of dimension $d_{\xi} = \dim(\dn{\Pi})-\dim( \vec{\dn{\lambda}})$;
 in our implementation  we have used as $\dn{\xi}$ the last $d_{\xi}$ terms of $\dn{\Pi}_{\xi}$.

\item Accept the proposed move with probability $\alpha = \min ( 1, A)$ with
            \begin{eqnarray}
            A &=& \frac{f\Big( \dn{n} | {\mathbf \wp}\big(\dn{\lambda}\myprime \big) \Big)
						            f \big(\vec{\dn{\lambda}}\myprime\big)
												f\left(\dn{\xi}' \right)
                        q \big(\vec{\dn{\lambda}}\myt, \, \dn{\xi}^{(t)} \big| \vec{\dn{\lambda}}\myprime, \, \dn{\xi}'\big) }
                       {f\Big( \dn{n} | {\mathbf \wp}\big(\dn{\lambda}^{(t)}\big) \Big)
											  f \big(\vec{\dn{\lambda}}\myt \big)
												f(\dn{\xi}^{(t)})
												q \big(\vec{\dn{\lambda}}\myprime, \, \dn{\xi}' \big|\vec{\dn{\lambda}}\myt, \, \dn{\xi}^{(t)} \big) }
                       \nonumber \\
              &=& \frac{f( \dn{n}|\dn{\Pi}'      ) f\big(\vec{\dn{\lambda}}\myprime \big)  f(\dn{\xi}') q(\dn{\Pi}^{(t)}  | \dn{\Pi}' ) }
                       {f( \dn{n}|\dn{\Pi}^{(t)} ) f\big(\vec{\dn{\lambda}}\myt \big)  f(\dn{\xi}) q(\dn{\Pi}' | \dn{\Pi}^{(t)}  ) }
                       \times
                       \mbox{abs} \left(
                       \frac{ {\cal J}\Big(\dn{\Pi}^{(t)}, \vec{\dn{\lambda}}\myt, \dn{\xi}^{(t)} \Big)}
                            { {\cal J}\Big(\dn{\Pi}', \vec{\dn{\lambda}}\myprime, \dn{\xi}' \Big)}
                       \right),
                                   \label{MH2_acc_prob}
            \end{eqnarray}
        where $\mbox{abs}(\cdot)$ stands for the absolute value,
        $\dn{\Pi}_{\xi}=\dn{\xi}$, and
        ${\cal J}={\cal J}( \dn{\Pi}, \vec{\dn{\lambda}}, \dn{\xi} ) $ is the determinant of the Jacobian matrix of the transformation
          $\dn{\Pi}=g( \vec{\dn{\lambda}}, \dn{\xi} )$
        specified by Equations  (\ref{joint_probs}), (\ref{augm_joint_probs}), and (\ref{lambda_of_marginal_model}).
                Similarly to Step 1, $\vec{\dn{\lambda}}\myt$, $\dn{\xi}^{(t)}$ and $\dn{\lambda}^{(t)}$
                are used to denote the values of the corresponding parameters in the current iteration $t$ of the algorithm.

\item If the move is accepted, then set $\dn{\Pi}^{(t+1)}=\dn{\Pi}'$,
      $\dn{\xi}^{(t+1)}=\dn{\xi}'$, and $\vec{\dn{\lambda}}^{(t+1)}=\vec{\dn{\lambda}}\myprime$
      otherwise set $\dn{\Pi}^{(t+1)}=\dn{\Pi}^{(t)}$ and $\vec{\dn{\lambda}}^{(t+1)}=\vec{\dn{\lambda}}^{(t)}$.
\end{enumerate}

The pseudo-parameter vector $\dn{\xi}$ is used only to retain the
dimension balance between the marginal log-linear parameterisation
and the probability parameterisation used in \cite{NtzoufrasTarantola2013}. Furthermore, it is directly matched to specific
probability parameters of the bi-directed graph $G$. Hence, we can
indirectly ``eliminate''   the effects of $\dn{\xi}$  on the algorithm by assuming
that its elements are uniformly distributed in the zero-one
interval. Under this view, we set $ f(\xi_\imath) = I_{ \{ 0 <
\xi_\imath < 1 \} } $ having as a result the elimination of the
ratio $f(\dn{\xi}')/f(\dn{\xi}^{(t)})$ from (\ref{MH2_acc_prob}).
In the following, we consider this choice in order to simplify all
proposed algorithms.

A good choice of the proposal $ q(\dn{\Pi}' | \dn{\Pi}^{(t)}  )$ will lead to high (close to one) acceptance rates.
Therefore, an efficient proposal for the Metropolis Hastings scheme described in this section,
intuitively seems to be the following
            \begin{equation}
            q(\dn{\Pi}' | \dn{\Pi}^{(t)}  )
            = \sum _{n_{\cal A}} f_q \big( \dn{\Pi}' | \dn{n}^{\prime\cal A}\big)
              f   \big( \dn{n}^{\prime\cal A} | \dn{\Pi}^{(t)}, \dn{n} \big),
            \label{proposal_expression}
            \end{equation}
where $f\big( \dn{n}^{\cal A} | \dn{\Pi}, \dn{n} \big)$ is the
distribution of  counts $\dn{n}^{\cal A}$ given the observed
frequency table $\dn{n}$ and the probability parameter set
$\dn{\Pi}$ of the augmented table induced by ${\cal A}$; and
$f_q\big( \dn{\Pi} | \dn{n}^{\cal A}\big)$ is the conditional
posterior distribution of the probability parameter vector
$\dn{\Pi}$ given a proposed set of augmented data $\dn{n}^{{\cal A}}$.
Considering all possible configurations in
\eqref{proposal_expression} within each MCMC iteration is
cumbersome and time-consuming. One solution can be obtained by
using a random sub-sample of $\dn{n}^{{\cal A}}$.
Hence, we construct our MCMC by employing just one
realisation of $\dn{n}^{\cal A}$ which will play the role of
intermediate nuisance parameter that facilitates the construction of a sensible proposal distribution.
This approach corresponds to an MCMC scheme
with a target posterior distribution given by
\begin{eqnarray*} f(
{\dn{\lambda}}^{\cal A}, \dn{n}^{\cal A} | \dn{n} ) &\propto&
f\big( \dn{n} | {\mathbf \wp}({\dn{\lambda}}) \big) f( \vec{\dn{\lambda}} )  f(\dn{\xi}) f(\dn{n}^{\cal A} ).
\end{eqnarray*}
The parameter vector $\dn{n}^{\cal A}$ plays the role of (nuisance) augmented data
with $f(\dn{n}^{\cal A} )$ being a pseudo-prior.
 More precisely, we introduce $\dn{n}^{\cal A}$   in order to augment the target posterior and to  achieve a dimension balance between the two parameterizations: the marginal log-linear parameterisation and the probability based one.
This  trick assists us to retain the balance between the two parametrisations and it enables us to implement  standard MCMC methods.

In order to simplify the MCMC configuration,  we consider a uniform distribution as  pseudo-prior over  all possible configurations of
$\dn{n}^{\cal A}$.
Under this formulation, the acceptance
probability in the Metropolis-Hastings  is equal to $\alpha = \min
( 1, A)$ with $A$ given by
            \begin{eqnarray}
            A
            &=& \frac{f\big( \dn{n}|\dn{\Pi}' \big) f\big(\vec{\dn{\lambda}}\myprime\big)
                        f_q\big( \dn{\Pi}^{(t)} | \dn{n}^{{\cal A}(t)}\big)
												f\big( \dn{n}^{{\cal A}(t)} | \dn{\Pi}', \dn{n} \big) }
                       {f\big( \dn{n}|\dn{\Pi}^{(t)}  \big) f\big(\vec{\dn{\lambda}}\myt \big)
                        f_q\big( \dn{\Pi}' | \dn{n}^{\prime\cal A}\big) f\big( \dn{n}^{\prime\cal A} | \dn{\Pi}^{(t)}, \dn{n} \big)  }
                       \times
                       \mbox{abs} \left(
                       \frac{ {\cal J}\big(\dn{\Pi}^{(t)}, \vec{\dn{\lambda}}\myt, \dn{\xi}^{(t)} \big)}
                            { {\cal J}\big(\dn{\Pi}', \vec{\dn{\lambda}}\myprime, \dn{\xi}' \big)}
                       \right).
                       \label{acc_ratio_augm}
            \end{eqnarray}

In (\ref{acc_ratio_augm}), the probability functions
$f\big( \dn{n}^{\prime \cal A} | \dn{\Pi}, \dn{n} \big)$ and
$f\big( \dn{n}^{{\cal A}(t)} | \dn{\Pi}', \dn{n} \big) $
are readily available from the model construction and the likelihood representation
of the augmented table.
We only need to specify
$ f_q \big( \dn{\Pi}' | \dn{n}^{\prime\cal A}\big)$
which is the first component of (\ref{proposal_expression})
and it has the form of a posterior conditionally on the frequencies of the augmented table.
For this component, we can exploit
the conditional conjugate approach of \cite{NtzoufrasTarantola2013}.
In order to do so, we consider as a ``prior'' $f_q( \dn{\Pi}  )$  a product of Dirichlet distributions
in order to obtain a conjugate ``posterior'' distribution $f_q( \dn{\Pi}' | \dn{n}^{\prime\cal A})$.
Under this approach,
and by further considering that
$f\big( \dn{n}|\dn{\Pi} \big)  f\big( \dn{n}^{\cal A}|\dn{\Pi}, \dn{n} \big)
= f\big( \dn{n}^{\cal A}|\dn{\Pi} \big)$,
then (\ref{acc_ratio_augm}) is further simplified to
            \begin{eqnarray}
            A&=& \frac{f\big( \dn{n}^{{\cal A}(t)} |\dn{\Pi}^{\prime} \big)f\big(\vec{\dn{\lambda}}\myprime\big)
                        f_q\big( \dn{\Pi}^{(t)} | \dn{n}^{{\cal A}(t)} \big)  }
                       {f\big( \dn{n}^{\prime\cal A}|\dn{\Pi}^{(t)}  \big) f\big(\vec{\dn{\lambda}}\myt \big)
                        f_q\big( \dn{\Pi}' | \dn{n}^{\prime\cal A}\big)    }
                       \times
                       \mbox{abs} \left(
                       \frac{ {\cal J}\big(\dn{\Pi}^{(t)}, \vec{\dn{\lambda}}\myt, \dn{\xi}^{(t)} \big)}
                            { {\cal J}\big(\dn{\Pi}', \vec{\dn{\lambda}}\myprime, \dn{\xi}' \big)}
                       \right) \,.\label{prior_based_alg}
            \end{eqnarray}
In the following of this manuscript, we will refer to this approach as the {\it probability-based independence sampler} (PBIS).

\subsection{Prior Adjustment Algorithm}
\label{section_paa}

As we have already stated, although PBIS simplifies the MCMC scheme, the parameter space is still considerably extended
by considering the augmented frequency table $\dn{n}^{\cal A}$.
We can further simplify PBIS by using the following two-step procedure:
\begin{description}
\item[~~~~~Step 1:] 
Run the Gibbs sampler of \cite{NtzoufrasTarantola2013} to obtain a  sample from $\dn{\Pi}$ (see Section \ref{sec41} for more details).

\item[~~~~~Step 2:] Use the sample of step 1 (or sub-sample of it) as a proposal in the general
               Metropolis-Hastings algorithm  with acceptance rate (\ref{MH2_acc_prob}).
\end{description}

Since the sample of Step 1 is obtained from an MCMC algorithm, auto-correlation will be present.
A random (independent) sub-sample from the posterior distribution can be obtained by following different strategies.  The most common approach can be obtained by using thinning, where we  keep one observation for every set of $K$  iterations. The thinning interval $K$ can be easily defined by monitoring autocorrelation function or using the
convergence diagnostic of \cite{raftery_lewis_1992}
which is available in {\sf R} packages such as {\sf CODA} or {\sf BOA}.
A drawback of this approach is that we may end up having a  considerably lower number of simulated observations.
For this reason, we suggest considering a random permutation of the full MCMC sample to ``destroy'' the induced autocorrelation.
We believe that the effect of this strategy will be minimal, since this sample is only used as a proposal in the  second MCMC procedure.
As a referee pointed out, by following either of these approaches,  the sample variance will get close to the independent variance asymptotically.
 Although intuitively this approach will lead to a sample from the target posterior distribution (or a close approximation of it), to our knowledge no mathematical proof is available.
 Indeed, empirical comparisons (see Section \ref{sim_study}) indicate no differences between the posterior distribution obtained by a correctly thinned MCMC sample and the re-ordered sample or even with the one-shot MCMC sampler of Section \ref{prior_adj}.
Therefore, we believe that  this sampler provides an efficient approximation of an independence Metropolis-Hastings algorithm.

Using such sample (or sub-sample) as proposed values within the  Metropolis-Hasting algorithm
is equivalent to using
the posterior distribution $f_q( \dn{\Pi} |  \dn{n} )$ as proposal in (\ref{MH2_acc_prob}),
that is
$q(\dn{\Pi}'  | \dn{\Pi}^{(t)} )  = f_q( \dn{\Pi}' |  \dn{n} ).$
Under this proposal,  (\ref{MH2_acc_prob})  now simplifies  to
            \begin{eqnarray}
            A &=& \frac{f(\vec{\dn{\lambda}}\myprime)          f_q( \dn{\Pi}^{(t)} )}
                        {f(\vec{\dn{\lambda}}\myt )    f_q( \dn{\Pi}' ) }
                       \times
                       \mbox{abs} \left(
                       \frac{ {\cal J}\Big(\dn{\Pi}^{(t)}, \vec{\dn{\lambda}}\myt, \dn{\xi}^{(t)} \Big)}
                            { {\cal J}\Big(\dn{\Pi}', \vec{\dn{\lambda}}\myprime, \dn{\xi}' \Big)}
                       \right).
                       \label{prior_adj_alg}
            \end{eqnarray}
We will refer to this algorithm as   the {\it prior-adjustment}
algorithm (PAA) due to its characteristic to correct for the
differences between the prior distributions used under the two
parameterisations.

The ``prior''  distribution $f_q(\dn{\Pi})$ is only used to build
the proposal. Therefore, it can be considered as a pseudo-prior.
It does not influence the target posterior distribution but only
affects the convergence rates of PAA. We can choose the parameters
of this pseudo-prior in such a way that (\ref{prior_adj_alg}) is
maximized and an optimal acceptance rate is achieved. When a
non-informative prior distribution for $\vec{\dn{\lambda}}$ is
used,  all Dirichlet parameters involved in $f_q(\dn{\Pi})$ can be set equal to one.
Under this choice,  the effect of the pseudo-prior is eliminated from the proposal, leaving the data-likelihood
to guide the MCMC algorithm.
PAA is less computationally demanding that the single-run MCMC algorithm introduced in Section \ref{prior_adj}, since
in we avoid four additional likelihood evaluations at each iteration required in the later.

\subsection{The Jacobian}

We conclude this section by providing analytical expressions of the Jacobian
required in the acceptance probabilities within each MCMC step in Sections \ref{prior_adj} and \ref{section_paa};
see Equations (\ref{prior_based_alg}) and (\ref{prior_adj_alg}).
        Specifically, the Jacobian terms  are given by
        ${\cal J}( \dn{\Pi}, \vec{\dn{\lambda}}, \dn{\xi} ) =\left| \dfrac{\partial \dn{\Pi}}{ \partial(\vec{\dn{\lambda}}, \dn{\xi})} \right|$
        and are simplified to
        \begin{eqnarray*}
        {\cal J}^{-1}
         =  \left| \dfrac{ \partial(\vec{\dn{\lambda}}, \dn{\xi})}{\partial \dn{\Pi}} \right|
         =   \left| \dfrac{ \partial(\vec{\dn{\lambda}}, \dn{\xi})}{\partial(\dn{\Pi}_{\xi}, \dn{\Pi}_{\setminus \xi})  }  \right|
         &=&  \left|
              \begin{matrix} \dfrac{\partial\vec{\dn{\lambda}}}{\partial  \dn{\Pi}_{\xi}  } &
                             \dfrac{\partial\vec{\dn{\lambda}}}{\partial  \dn{\Pi}_{\setminus \xi}  } \\
                             \dfrac{ \partial \dn{\xi}}{\partial \dn{\Pi}_{\xi} } &
                             \dfrac{ \partial \dn{\xi}}{\partial  \dn{\Pi}_{\setminus \xi}  }
              \end{matrix} \right| \\
          &=& \left| \begin{matrix} \dfrac{ \partial\vec{\dn{\lambda}}}{\partial \dn{\Pi}_{\xi}  } &
                                  \dfrac{ \partial\vec{\dn{\lambda}}}{\partial  \dn{\Pi}_{\setminus \xi}  }\\
                                  \dn{I} & \dn{0}
                   \end{matrix} \right|
           = -\left| \begin{matrix} \dfrac{ \partial\vec{\dn{\lambda}}}{\partial  \dn{\Pi}_{\setminus  \xi}  }
 \end{matrix} \right|,
\end{eqnarray*} where $\dn{\Pi}_{\setminus \xi}$ is obtained from $\dn{\Pi}$
 excluding the elements of $\dn{\Pi}_{\xi}$.

The elements of the Jacobian matrix are given by
\begin{equation}
\frac{\partial \lambda_k}{\partial  \Pi_\jmath}
=
\sum_{l=1}^{ \textsc{c}_{C} } \left\{ C_{k l} \left( \sum_{\imath=1}^{|{\cal I}|} M_{l \imath} P_\imath \right)^{-1}
              \sum \limits_ {\imath=1}^{|{\cal I}|} M_{l \imath} \Delta_{\imath \jmath} \right\}
              \mbox{~with~}
               \Delta_{\imath \jmath} = \frac{\partial P_\imath}{\partial \Pi_\jmath
               },
\label{jacelements}
\end{equation}
where ${P}_\imath$ denote the $\imath$ element of $\dn{P}$, and
$\textsc{c}_{C}$ is the number of columns of the contrast matrix
$\dn{C}$. For the saturated model,  the above equation simplifies
to
$$
\frac{\partial \lambda_k}{\partial P_\jmath }
= \sum_{l=1}^{ \textsc{c}_{C} }
  \frac{ C_{k l}  ( M_{l \jmath} - M_{l |{\cal I}|} ) }
       { \sum_{\imath=1}^{|{\cal I}|} M_{l \imath} P_\imath }
$$
since $\dn{\Pi}=\dn{P}$,  and
$P_{|{\cal I}|} = 1 - \sum_{\imath=1}^{|{\cal I}|-1} P_\imath$.
More details on the calculation of (\ref{jacelements}) are provided in Appendix \ref{app_jac}. 

In order to complete the specification of  (\ref{jacelements}), we
need to calculate the derivative terms $\Delta_{\imath \jmath}$.
Let us now denote by $\ivec$  the index of vector $\dn{P}$
such that $P_{\ivec} \equiv p( i )$.
Moreover,
the index $\jmath$ corresponds to a variable $u_\jmath \in {\cal A}$ such
that $\Pi_\jmath \equiv \pi_{u_\jmath | \pa(u_\jmath)} \big(j_{u_\jmath} | j_{\pa(u_\jmath)} \big)$
for a specific cell $j$ of the augmented table ${\cal I}^{\cal A}$.
Therefore, for any $\imath=\ivec$, terms $\Delta_{\imath \jmath}$ are given by
\be
\Delta_{\imath \jmath}=\Delta_{\ivec \jmath}=\frac{\partial P_\ivec}{\partial \Pi_\jmath}
= \frac{\partial p(i)}{\partial \pi_{u_\jmath | \pa(u_\jmath)}
\big( j_{u_\jmath} | j_{\pa(u_\jmath)} \big) } \mbox{~for every~}
\ivec \mapsto i \mbox{~and~} \jmath \mapsto (u_\jmath \, , j  ).
\label{delta}
\ee

For the computation of each $\Delta_{\imath \jmath}$ we consider two different cases:
(A)  $u_\jmath \in {\cal V}$ and (B) $u_\jmath \in {\cal L}$.
In the following, to simplify notation, we denote $u_\jmath$ by $u$.
Furthermore, we indicate with  ${\cal L}_u= {\cal L} \cap \pa(u)$ the latent variables that are parents of $u$, and with ${\cal
A}_u={\cal V} \cup \{ u \} \cup {\cal L}_u $.

For {\bf case A}, when $u$ is an observed variable, we obtain
\be
\dfrac{\partial  p(i) }
     {\partial \pi_{u|\paua} \big(j_u| j_{\paua} \big) } =
     \delta( i, j)
     \frac{  p^{ {\cal A}_u } \big( i, j_{ {\cal L}_u }   ) } { \pi_{u|\paua} ( i_u  |  j_{\paua} \big) }
\label{jac1a}
\ee
with
\be
\delta( i, j) =
\begin{cases}
~~~1  & \mbox{~if~} i_u = j_u<|{\cal I}_u|$ and $i_{\paua\setminus {\cal L}} = j_{\paua\setminus {\cal L}}\vspace{0.5em} \\
  -1  & \mbox{~if~} j_u \neq i_u = |{\cal I}_u| $ and $i_{\paua\setminus {\cal L}} = j_{\paua\setminus {\cal L}}\\
~~~0  &  \mbox{~if~} j_u \neq i_u < |{\cal I}_u| $ or
$\;\;i_{\paua\setminus L} \neq j_{\paua\setminus {\cal L}}
\end{cases}\, ,
\label{jac1b} \ee where 
$$
p^{ {\cal A}_u } \big( i, j_{ {\cal L}_u} \big) =
\begin{cases}
P \Big( Y_{\cal V} = i,  Y_{{\cal L} \cap \pa(u)} = j_{{\cal L} \cap \pa(u)} \Big) & {\cal L}_u \neq \emptyset \\
p(i) & {\cal L}_u =\emptyset
\end{cases}.
$$

For {\bf case B}, when $u$ is a latent variable, $\pa(u) = \emptyset$
due to the structure of the DAG representation.
Hence, the derivative is  given by \be \dfrac{\partial  p(i) } {\partial
\pi_{u|\paua} \big(j_u| j_{\paua} \big) } = \dfrac{\partial  p(i)
} {\partial \pi_{u} \big(j_u \big) } =  \dfrac{ p^{{\cal A}_u}( i,
j_u ) }{ \pi_u(j_u)}
  -\dfrac{ p^{{\cal A}_u}\big( i, |{\cal I}_u| \big) }{ \pi_u\big( |{\cal I}_u|\big) }
  \mbox{~for~} j_u < |{\cal I}_u|~.
\label{jac2}
\ee

Detailed derivation of expressions (\ref{jac1a})--(\ref{jac2}) are available in Appendix \ref{app_jac}. 

%

\section{Illustrative Examples}
\label{Examples}

\subsection{Simulation Study}

\label{sim_study}

In this section, we evaluate the performance of the proposed methodology via a simulation study.
We  generated $100$ samples from the marginal association model represented by the  bi-directed graph  of Figure \ref{bidir} and true log-linear
 interactions given  in  Table \ref{true_effects}.
This model encodes the marginal independencies  $\jind{Y_{\{A,B\}}}{Y_{D}}$ and  $\jind{Y_{A}}{Y_{\{D,C\}}}$ under the  connected set Markov property.

\begin{table}[h]
\caption{True effect values used for the simulation study} 
\label{true_effects}
\[
\begin{array}{lll}
\hline
\mbox{Marginal} & \mbox{Active interactions} & \mbox{Zero interactions} \\
\hline
\mbox{AC} & \lambda^{AC}_\emptyset =-1.40, \lambda^{AC}_A(2) =-0.15,  \lambda^{AC}_C(2) =0.10, & \lambda^{AC}_{AC}=0\\
\mbox{AD} &\lambda^{AD}_B(2)       = 0.12,     &\lambda^{BD}_{BD}(2,2)     =0     \\
\mbox{BD} &\lambda^{BD}_D(2)       =-0.09,     &\lambda^{AD}_{AD}(2,2)     = 0    \\
\mbox{ACD}&\lambda^{ACD}_{CD}(2,2)     =0.20,  &\lambda^{ACD}_{ACD}(2,2,2) =0     \\
\mbox{ABD}&\lambda^{ABD}_{AB}(2,2)     =-0.15, &\lambda^{ABD}_{ABD}(2,2,2) =0     \\
\mbox{ABCD}&\lambda^{ABCD}_{BC}(2,2)   =-0.30, \lambda^{ABCD}_{ABC}(2,2,2) =0.15, \\
                     &\lambda^{ABCD}_{BCD}(2,2,2)=-0.10, \lambda^{ABCD}_{ABCD}(2,2,2)=0.07. \\
\hline
\end{array}
\]
\end{table}

We compare the methods introduced and discussed in this article  in terms of acceptance rate, effective sample (ESS)  per second of CPU time and the Monte Carlo Error (MCE).
In addition to
 the algorithms described in Section \ref{our_mcmc} (PBIS and PAA) we also consider  random walks on marginal log-linear interactions $\lambda$ and on logits of probability parameters $\pi$ (RW-$\lambda$ and RW-$\pi$ respectively).
For all interactions and for each method under consideration,  we calculated the ESS per second of CPU time  using
both  {\sf coda} and {\sf rstan} packages in {\sf R}.

We calculated   MCEs   for the mean and standard deviation of all interactions via the batch mean method using $50$ batches of equal size.
The previous quantities have been adjusted for computational time by fixing the number of iterations of PAA
and then considering as number of iterations for the remaining methods the ones which correspond to the computational time of PAA.
\clearpage

In order to proceed with a more rigorous analysis, we first present  the results for a single randomly selected sample (see Table \ref{simex_data} for the specific dataset under consideration)
and then we discuss the main important  findings  for all samples.

\begin{table}[hbt]\caption{Simulated dataset} \label{simex_data}
\begin{center}
\begin{tabular}{ ccc c cc c cc c cc}
\hline
  & \multicolumn{5}{c}{ {\it $D_1$ }} && \multicolumn{5}{c}{ {\it $D_2$ }} \\
    \cline{2-6}\cline{8-12}
  & \multicolumn{2}{c}{ {\it $C_1$ }} && \multicolumn{2}{c}{ {\it $C_2$ }}&& \multicolumn{2}{c}{ {\it $C_1$ }}&& \multicolumn{2}{c}{ {\it $C_2$ }}\\
    \cline{2-3}\cline{5-6}\cline{8-9}\cline{11-12}
  A & ($B_1$) & ($B_2$) && ($B_1$) & ($B_2$)&& ($B_1$) & ($B_2$)&& ($B_1$) & ($B_2$)\\
\hline
  $A_1$ &    25   & 44 &&    47   & 21 &&     6   & 36 &&    65   & 29 \\
  $A_2$ &    31   & 25 &&    31   & 12 &&    27   & 17 &&    65   & 19 \\
\hline
\end{tabular}
\end{center}
\vspace{-2em}
\end{table}

 In terms of acceptance rate, PAA achieves a $50\%$ rate which is
satisfactory for an independence sampler and  considerably higher than the corresponding one achieved by the
PBIS algorithm ($\approx 15\%$).
This could be justified by the fact that, in the later method, increased uncertainty was introduced by the data augmentation approach,
which expands the parameter space by  introducing the latent counts as proposed in Section   \ref{prior_adj}.
On the other hand,  the RW algorithms (either for $\lambda$ or for $\pi$) were tuned
in order to achieve an acceptance rate close to $35\%$.
 The acceptance rates of the independence samplers (PBIS and PAA) and of the  RW algorithms cannot be compared due to
the different nature of the proposals.

\begin{figure}[t!]
      \caption{ESS per   second of CPU   time  for the simulated data of Table \ref{simex_data}}
            \label{plot_acc_simex}
      \vspace{-4em}
      \begin{center}
      \psfrag{RW-pi}[l][c][0.82]{\hspace{-1em}\sffamily RW-$\pi$  }
      \psfrag{RW-lambda}[l][c][0.82]{\hspace{-3em} \sffamily RW-$\lambda$}

      \includegraphics[width=14cm]{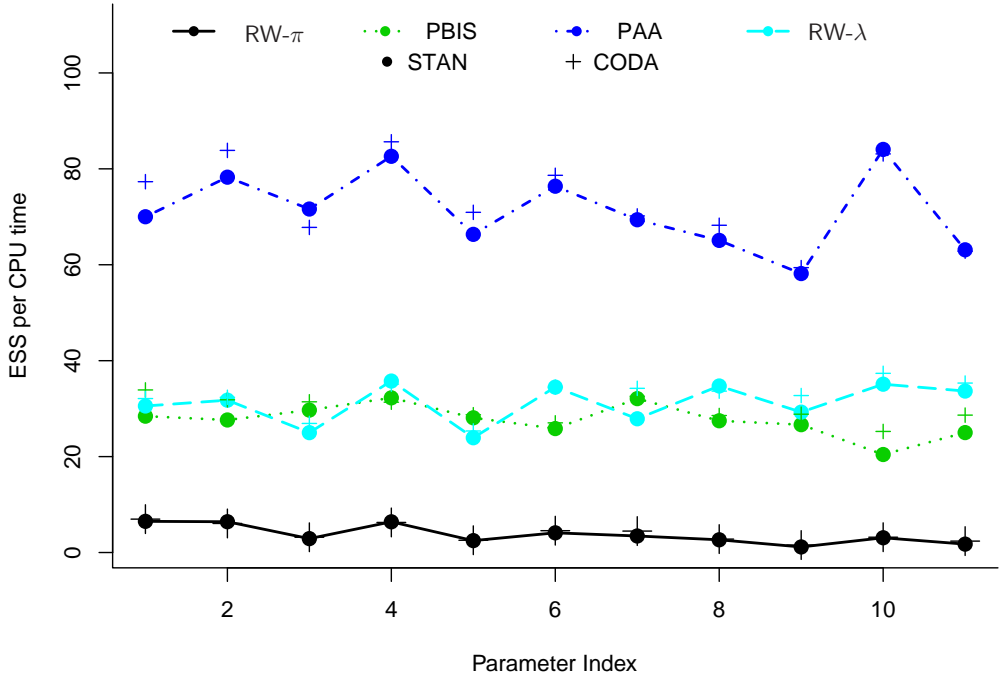}
      \end{center}
			\vspace{-2em}
\end{figure}

Summary statistics of ESS per second of CPU time are presented in Table \ref{ESSpt_simex},
while the detailed ESS for all interactions are depicted in Figure \ref{plot_acc_simex}.
 These  results indicate that PAA is the most efficient method followed by PBIS and RW-$\lambda$ having similar values,
while  the RW-$\pi$ appears as a rather inefficient way to approach the problem.

\begin{table}[hbp]
\caption{Summary statistics of ESS per second of CPU time for the simulated data of Table \ref{simex_data} }
\label{ESSpt_simex}
\begin{center}
\begin{tabular}{l | ccccc@{~~}c@{~~}ccccc}
\hline
\hline
          &\multicolumn{5}{c}{CODA} &&\multicolumn{5}{c}{STAN}  \\
                    \cline{2-6} \cline{8-12}
Method    & Min   &  $Q_1$& Median & $Q_3$ & Max  && Min   &  $Q_1$& Median & $Q_3$ & Max   \\
\hline
RW  on $\pi$     & 1.6   & 2.7   & 3.2   & 5.3  & 7.0  && 1.2  & 2.6  & 3.1 & 5.3   & 6.5 \\
RW  on $\lambda$ & 25.3  & 32.2  & 33.7  & 34.8 & 37.4 && 23.9 & 28.6 & 31.8 & 34.6 & 35.8\\
PBIS             & 25.2  & 28.6  & 28.8  & 31.6 & 33.9 && 20.4 & 26.3 & 27.6 & 29.1 & 32.3\\
PAA              & 59.4  & 68.0  & 70.9  & 80.9 & 85.6 && 58.2 & 65.7 & 70.0 & 77.3 & 84.1\\
\hline
\multicolumn{12}{p{13cm}}{ \footnotesize\it Min, $Q_1$, Median, $Q_3$, Max: Minimun, first quantile, Median,  third quantile and maximun of acceptance rates over different interactions} \\[-0.7em]
\end{tabular}
\end{center}
\vspace{-1em}
\end{table}

\begin{figure}[!p]
      \caption{Posterior densities for each parameter of the chain model estimated for the simulated data of Table \ref{simex_data}}
      \label{posterior_simex}
            \vspace{-1em}
            \begin{center}

      \psfrag{A2}[c][b][0.7]{$\lambda_{\scriptscriptstyle A}^{\scriptscriptstyle AC}$}
      \psfrag{C2}[c][b][0.7]{$\lambda_{\scriptscriptstyle C}^{\scriptscriptstyle AC}$}
      \psfrag{B2}[c][b][0.7]{$\lambda_{\scriptscriptstyle B}^{\scriptscriptstyle BD}$}
      \psfrag{D2}[c][b][0.7]{$\lambda_{\scriptscriptstyle D}^{\scriptscriptstyle AD}$}

      \psfrag{C2D2}[c][b][0.7]{$\lambda_{\scriptscriptstyle C D}^{\scriptscriptstyle ACD}$}
            \psfrag{A2B2}[c][b][0.7]{$\lambda_{\scriptscriptstyle A B}^{\scriptscriptstyle ABD}$}

      \psfrag{B2C2}[c][b][0.7]{$\lambda_{\scriptscriptstyle B C}^{\scriptscriptstyle ABCD}$~}
      \psfrag{A2B2C2}[c][b][0.7]{$\lambda_{\scriptscriptstyle A B C}^{\scriptscriptstyle ABCD}$}
      \psfrag{B2C2D2}[c][b][0.7]{~~$\lambda_{\scriptscriptstyle B C D}^{\scriptscriptstyle ABCD}$}
            \psfrag{A2B2C2D2}[c][b][0.7]{~~~~~~~~$\lambda_{\scriptscriptstyle ABCD}^{\scriptscriptstyle ABCD}$}

      \psfrag{RW-pi}[l][c][0.45]{\hspace{-1.8em}\sffamily RW-$\pi$  }
      \psfrag{RW-lambda}[l][c][0.45]{\hspace{-3.4em} \sffamily RW-$\lambda$}

      \includegraphics[width=12cm]{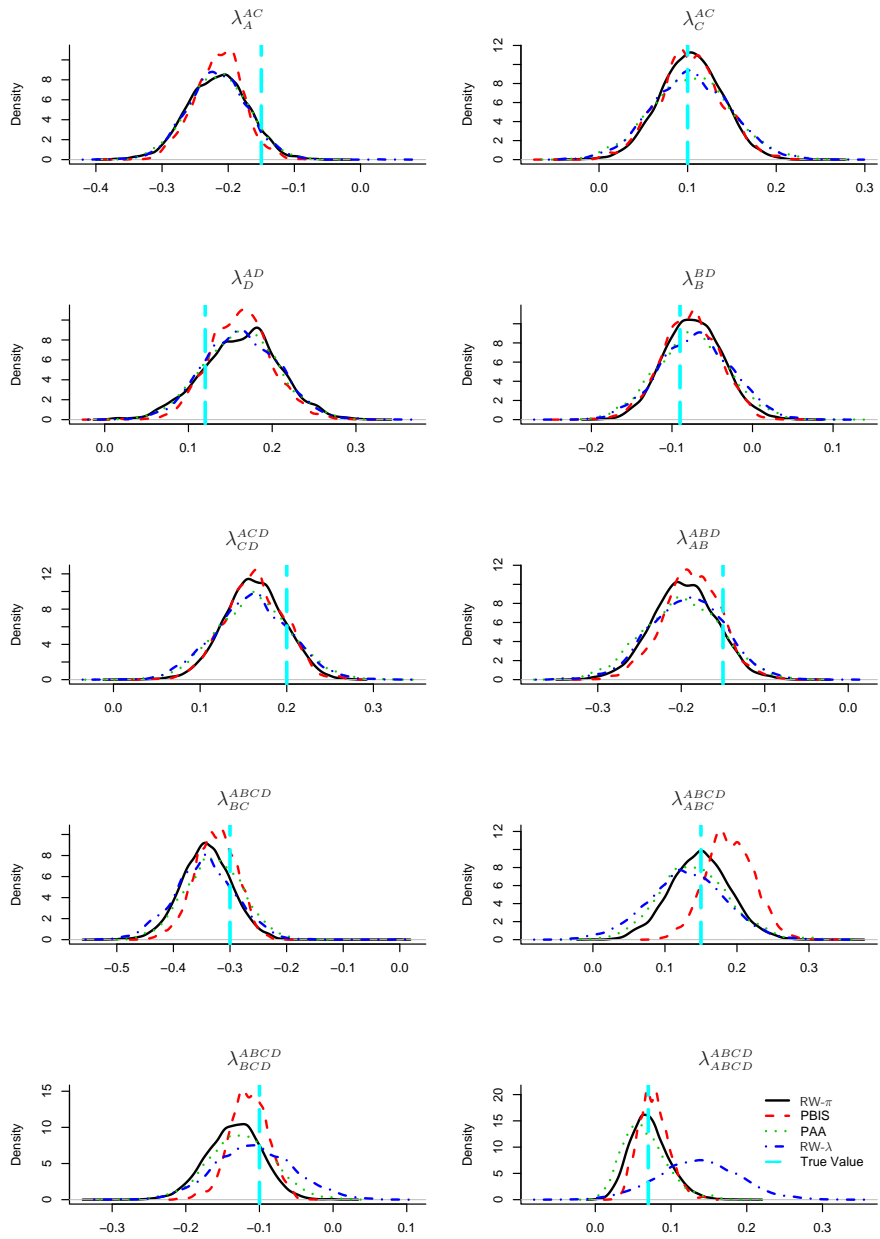}
      \end{center}
\end{figure}

Figure \ref{posterior_simex} presents the estimated posterior distributions for all interactions.
No major differences are observed in most interactions with  the true  values being in the
centre (or near the centre) of the posterior distribution of interest.
Some differences are observed in higher order interaction terms
$\lambda^{ABCD}_{ABC}$, $\lambda^{ABCD}_{BCD}$ and $\lambda^{ABCD}_{ABCD}$
estimated from the $ABCD$ marginal table.
More specifically, for $\lambda^{ABCD}_{ABC}$ only  RW-$\pi$ seems to estimate exactly the true value while PBIS slightly overestimates this effect
and the rest of the methods underestimate it.
  For  $\lambda^{ABCD}_{BCD}$,   all methods provide similar posterior distributions with the PBIS having lower dispersion.
Finally, for the four-way interaction, all methods correctly identify the true effect   except for
RW-$\lambda$  that overestimates the true value.
Generally, the estimated posterior distribution obtained by our proposed method, PAA, seems that correctly identifies the true  values for all interactions.

\begin{figure}[!h]
      \caption{MCEs for posterior mean adjusted for time for the simulated data of  of Table \ref{simex_data}}
            \label{MCEs_adj_means}
            \vspace{-1em}
            \begin{center}
        \psfrag{RW-pi}[Bc][Bc][0.60]{\sffamily RW-$\pi$}
        \psfrag{RW-lambda}[Bc][Bc][0.6]{\hspace{-3.4em} \sffamily RW-$\lambda$}
      \includegraphics[trim=0 0 0 1.5cm, clip=true, width=11cm]{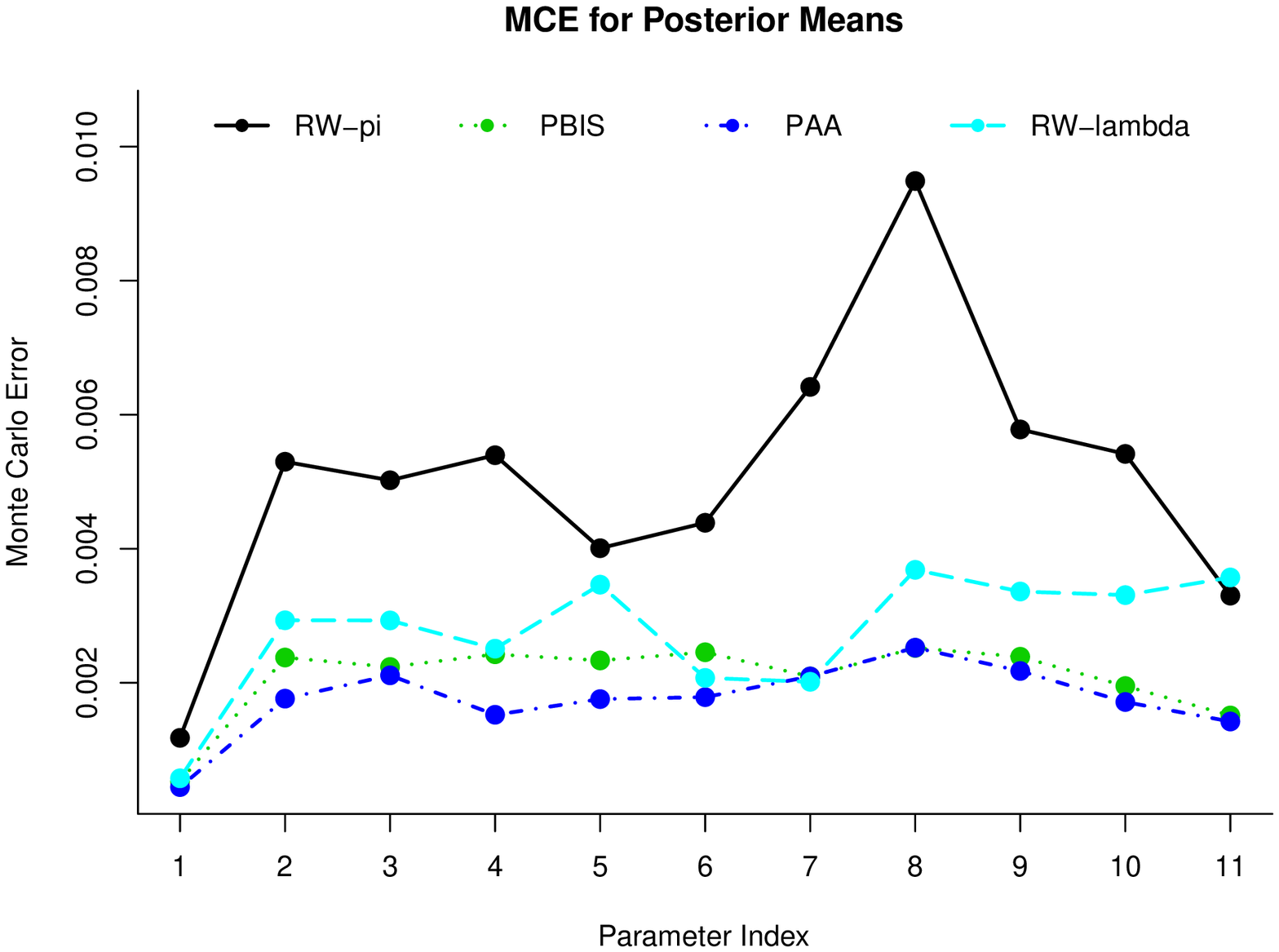}
      \end{center}
            \vspace{-3em}
\end{figure}

In Figure \ref{MCEs_adj_means} we represent   the time adjusted MCEs for the posterior means
for all methods and all marginal log-linear interactions.
We notice that PAA performs better than all competing methods, since the corresponding MCEs are lower  for almost all interactions.
In contrast, more dispersion is observed for the posterior distribution of  RW-$\pi$ that is clearly performing worse than the other methods.

\begin{figure}[!h]
      \caption{MCEs for posterior standard deviations  adjusted for CPU time for the simulated data of  Table \ref{simex_data}}
            \label{MCEs_adj_sds}
            \vspace{-1em}
            \begin{center}
        \psfrag{RW-pi}[Bc][Bc][0.60]{\sffamily RW-$\pi$}
        \psfrag{RW-lambda}[Bc][Bc][0.6]{\hspace{-2em} \sffamily RW-$\lambda$}
     \includegraphics[trim=0 0 0 1.5cm, clip=true,width=11cm]{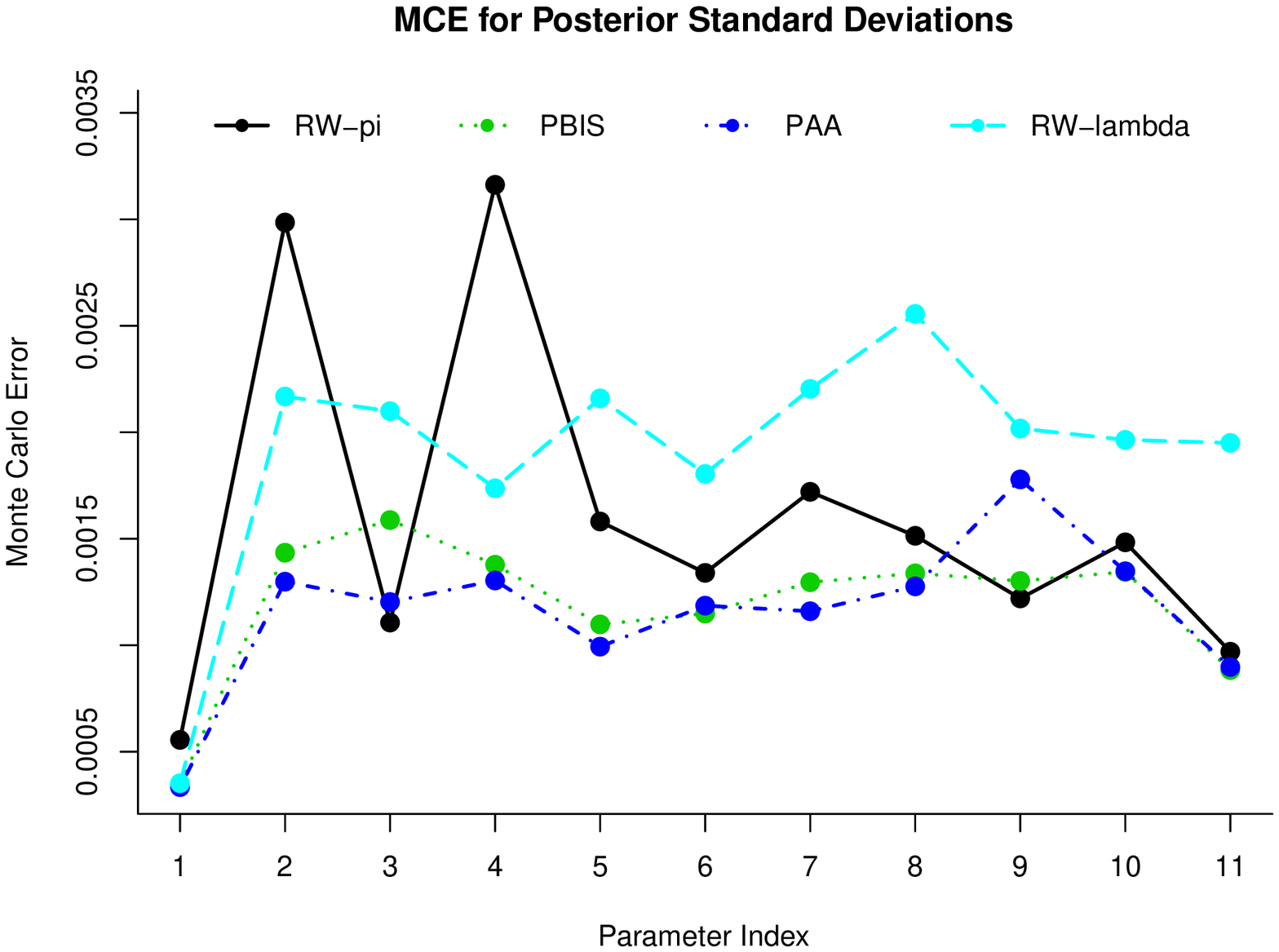}
      \end{center}
            \vspace{-3em}
\end{figure}

Regarding the posterior standard deviations, Figure \ref{MCEs_adj_sds} depicts time adjusted MCEs
for all marginal log-linear interactions under all methods under consideration.
PAA and PBIS demonstrate overall a better performance (in terms of Monte Carlo variability) in comparison to RW-$\lambda$ and RW-$\pi$.

\begin{figure}[hbp]
      \caption{ESS per   second of CPU   time over 100 simulated datasets}
            \label{sim_study_ess}
            \vspace{-1em}
            \begin{center}

      \psfrag{A2}[c][b][0.7]{$\lambda_{\scriptscriptstyle A}^{\scriptscriptstyle AC}$}
      \psfrag{C2}[c][b][0.7]{$\lambda_{\scriptscriptstyle C}^{\scriptscriptstyle AC}$}
      \psfrag{B2}[c][b][0.7]{$\lambda_{\scriptscriptstyle B}^{\scriptscriptstyle BD}$}
      \psfrag{D2}[c][b][0.7]{$\lambda_{\scriptscriptstyle D}^{\scriptscriptstyle AD}$}

      \psfrag{C2D2}[c][b][0.7]{$\lambda_{\scriptscriptstyle C D}^{\scriptscriptstyle ACD}$}
            \psfrag{A2B2}[c][b][0.7]{$\lambda_{\scriptscriptstyle A B}^{\scriptscriptstyle ABD}$}

      \psfrag{B2C2}[c][b][0.7]{$\lambda_{\scriptscriptstyle B C}^{\scriptscriptstyle ABCD}$~}
      \psfrag{A2B2C2}[c][b][0.7]{$\lambda_{\scriptscriptstyle A B C}^{\scriptscriptstyle ABCD}$}
      \psfrag{B2C2D2}[c][b][0.7]{~~$\lambda_{\scriptscriptstyle B C D}^{\scriptscriptstyle ABCD}$}
            \psfrag{A2B2C2D2}[c][b][0.7]{~~~~~~~~$\lambda_{\scriptscriptstyle ABCD}^{\scriptscriptstyle ABCD}$}

        \psfrag{RW-pi}[Bc][Bc][0.45]{\sffamily RW-$\pi$}
        \psfrag{RW-l}[Bc][Bc][0.45]{\sffamily RW-$\lambda$}

      \includegraphics[width=11cm]{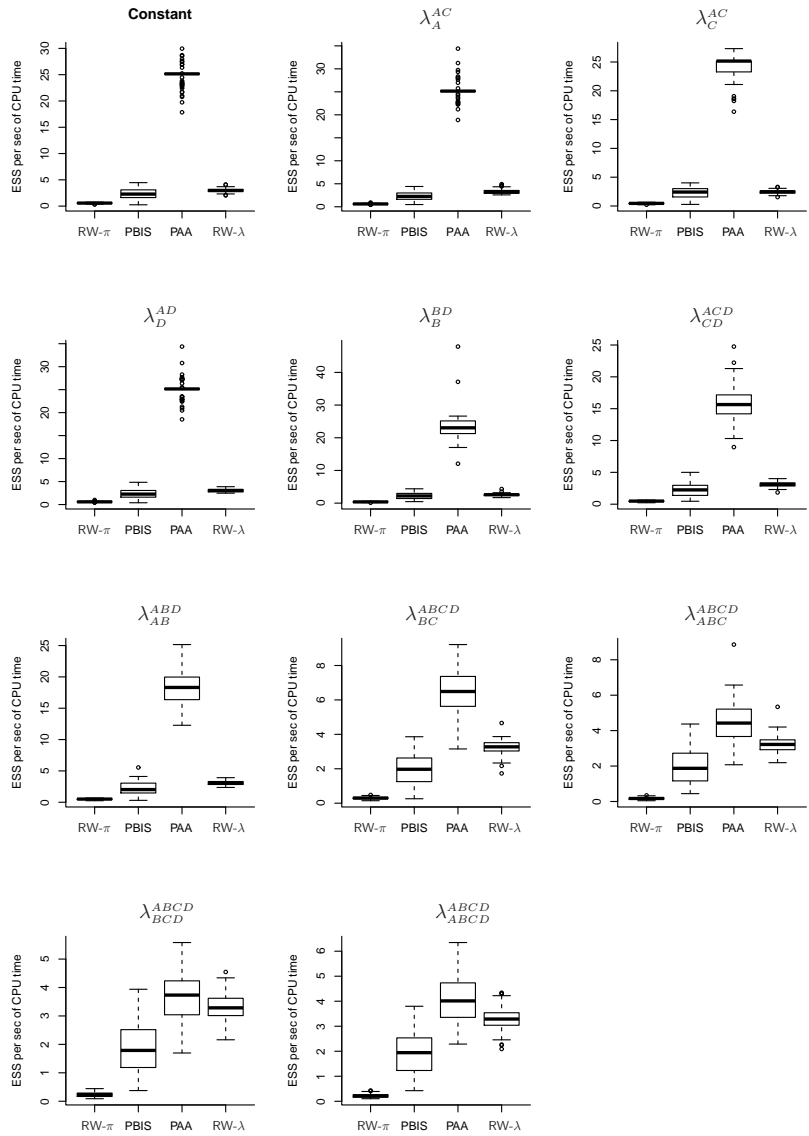}
      \end{center}
            \vspace{-2em}
\end{figure}

\begin{figure}[!h]
      \caption{Posterior means of marginal log-linear interactions over 100 simulated datasets}
            \label{sim_study_post}
            \vspace{-1em}
            \begin{center}

      \psfrag{A2}[c][b][0.7]{$\lambda_{\scriptscriptstyle A}^{\scriptscriptstyle AC}$}
      \psfrag{C2}[c][b][0.7]{$\lambda_{\scriptscriptstyle C}^{\scriptscriptstyle AC}$}
      \psfrag{B2}[c][b][0.7]{$\lambda_{\scriptscriptstyle B}^{\scriptscriptstyle BD}$}
      \psfrag{D2}[c][b][0.7]{$\lambda_{\scriptscriptstyle D}^{\scriptscriptstyle AD}$}

      \psfrag{C2D2}[c][b][0.7]{$\lambda_{\scriptscriptstyle C D}^{\scriptscriptstyle ACD}$}
            \psfrag{A2B2}[c][b][0.7]{$\lambda_{\scriptscriptstyle A B}^{\scriptscriptstyle ABD}$}

      \psfrag{B2C2}[c][b][0.7]{$\lambda_{\scriptscriptstyle B C}^{\scriptscriptstyle ABCD}$~}
      \psfrag{A2B2C2}[c][b][0.7]{$\lambda_{\scriptscriptstyle A B C}^{\scriptscriptstyle ABCD}$}
      \psfrag{B2C2D2}[c][b][0.7]{~~$\lambda_{\scriptscriptstyle B C D}^{\scriptscriptstyle ABCD}$}
            \psfrag{A2B2C2D2}[c][b][0.7]{~~~~~~~~$\lambda_{\scriptscriptstyle ABCD}^{\scriptscriptstyle ABCD}$}

                \psfrag{RW-pi}[Bc][Bc][0.5]{\sffamily   RW-$\pi$  }
        \psfrag{RW-l}[Bc][Bc][0.5]{\hspace{0.2cm} \sffamily RW-$\lambda$}

      \includegraphics[width=14cm]{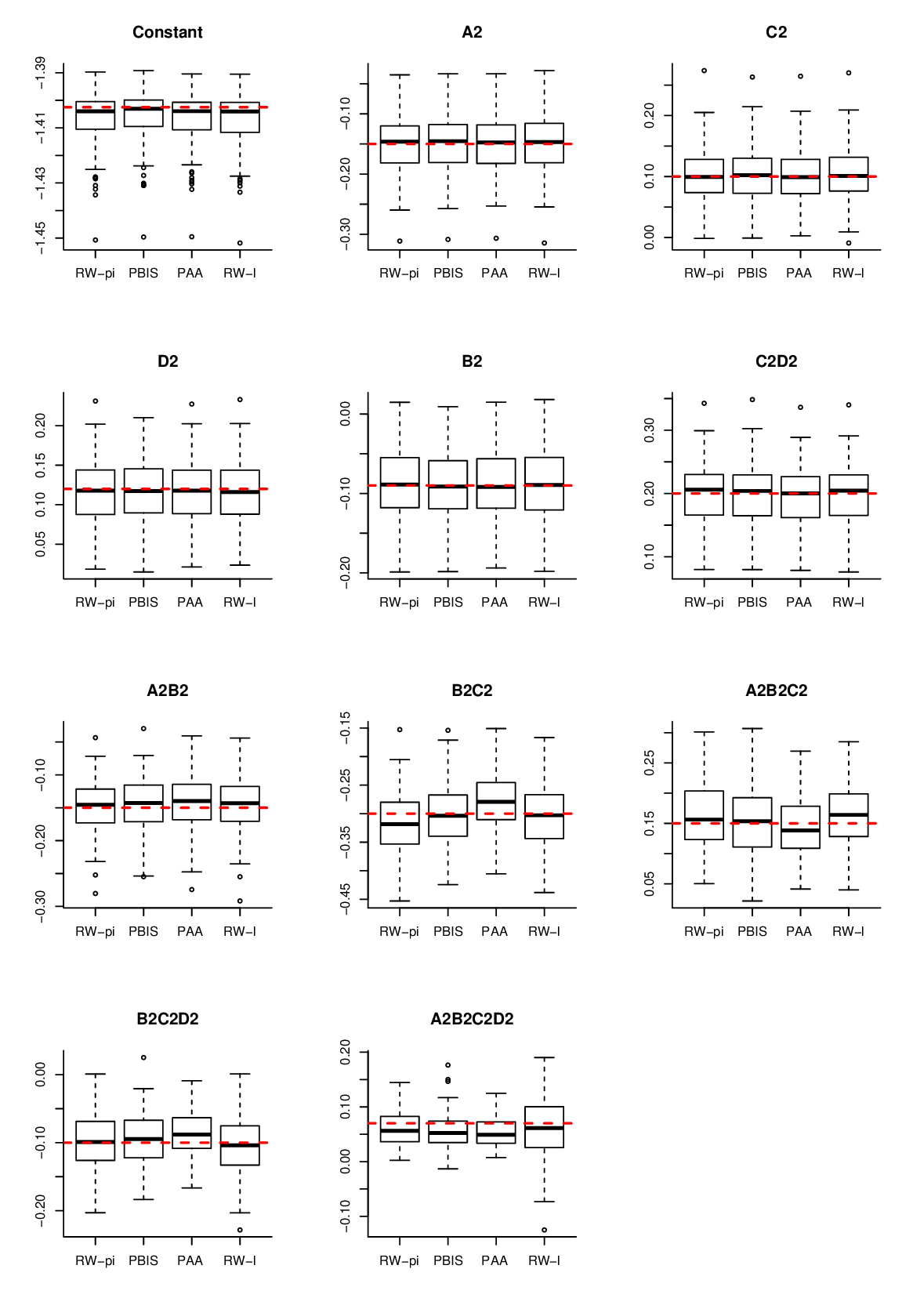}
        \end{center}
\end{figure}

Regarding the simulation study,
we present the ESS per second of CPU time and the posterior mean for each parameter
over 100 generated datasets
in Figures \ref{sim_study_ess} and \ref{sim_study_post}, respectively.
From the distribution of the ESS per  second, we confirm the results found in the single-sample analysis
which indicate that PAA is clearly the most efficient between the four methods under consideration.
PBIS and the RW on $\lambda$ are equally efficient in terms of ESS/minute
while the RW-$\pi$ is the least efficient method.
>From Figure \ref{sim_study_post} we confirm that the estimated posterior means under all  methods
successfully identify (with minor deviances) the true parameter.

\begin{figure}[hbp]
      \caption{MCEs for posterior Mean adjusted for CPU time for the 100 datasets of the simulation study}
            \label{MCEs_adj_means100}
            \vspace{-1em}
            \begin{center}
        \psfrag{RW-pi}[Bc][Bc][0.60]{\sffamily RW-$\pi$}
        \psfrag{RW-l}[Bc][Bc][0.6]{\hspace{-2em} \sffamily RW-$\lambda$}
     \includegraphics[trim=0 0 0 1.5cm, clip=true,width=11cm]{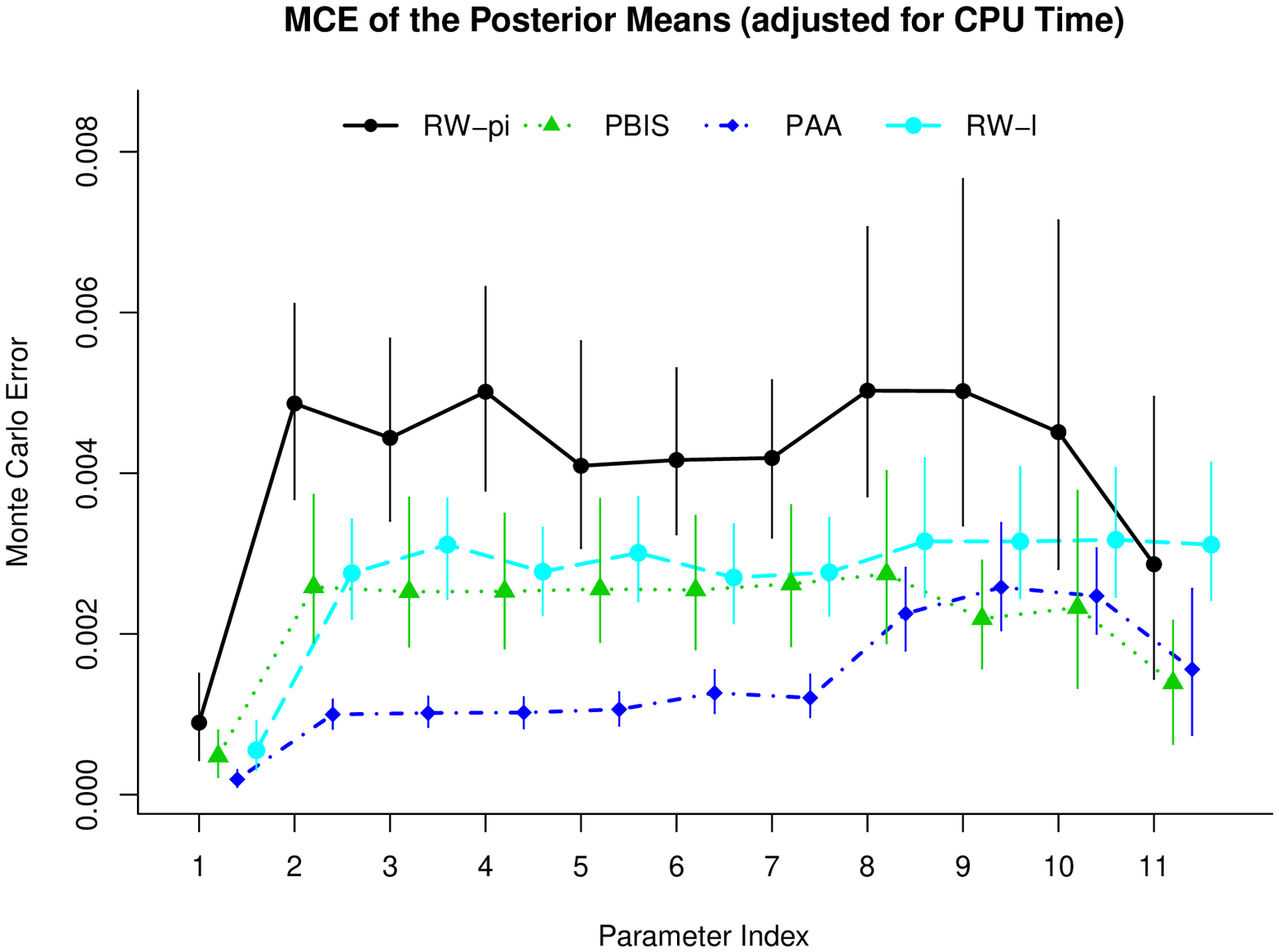}
      \end{center}
            \vspace{-3em}
\end{figure}

\begin{figure}[hbp]
      \caption{MCEs for posterior Standard Deviations  adjusted for CPU time for the 100 datasets of the simulation study}
            \label{MCEs_adj_sds100}
            \vspace{-1em}
            \begin{center}
        \psfrag{RW-pi}[Bc][Bc][0.60]{\sffamily RW-$\pi$}
        \psfrag{RW-l}[Bc][Bc][0.6]{\hspace{-2em} \sffamily RW-$\lambda$}
     \includegraphics[trim=0 0 0 1.5cm, clip=true,width=11cm]{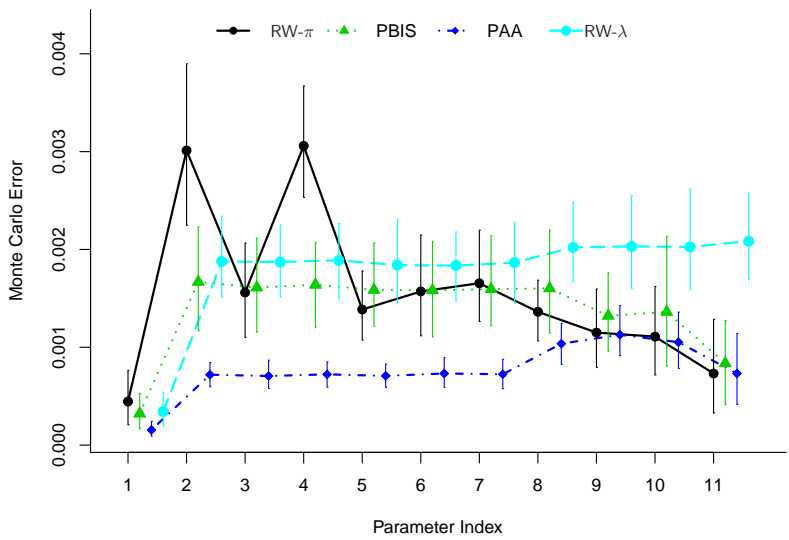}
      \end{center}
            \vspace{-3em}
\end{figure}

Figures \ref{MCEs_adj_means100} and \ref{MCEs_adj_sds100} present the $95 \%$ error bars of  the average time adjusted MCEs   for the  posterior means and standard deviations.
Each error bar represents the $2.5$ and the $97.5$ percentiles as well as the average of the quantities of interest (posterior means and standard deviations) for every interaction
across all generated data sets.
In most cases, PAA achieves values lower or at least of comparable size to the corresponding one of the  other methods.

\clearpage

\begin{figure}[h]
      \caption{Boxplots of the ratio of posterior standard deviations of PAA vs. RW-$\lambda$ for the simulated 100 datasets of the simulation study}
            \label{sd_ratios}
            \vspace{-1em}
            \begin{center}

      \psfrag{YLAB}[Bc][Bc][0.60]{\sf Posterior SD-Ratio (PAA vs. RW-$\lambda$) }
      \includegraphics[width=11cm]{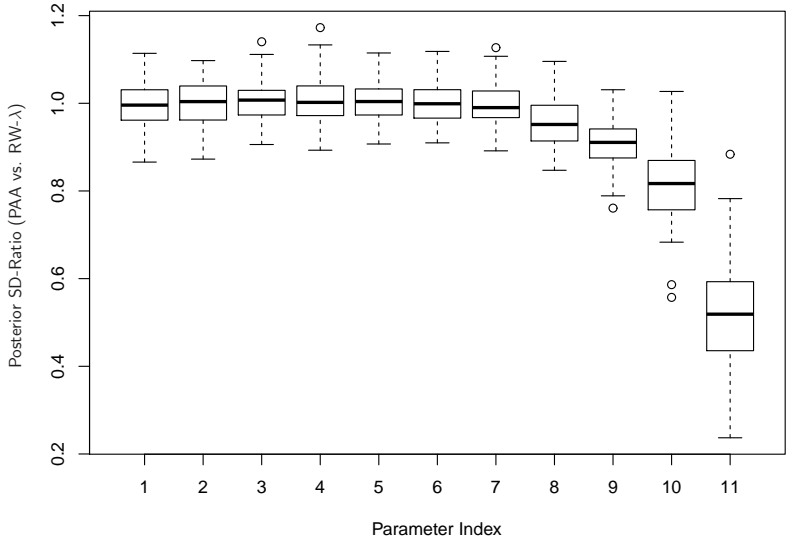}
      \end{center}
            \vspace{-3em}
\end{figure}

We conclude this section with a comparison of the dispersion of  PAA and RW-$\lambda$.
Figure \ref{sd_ratios} illustrates the distributions of the ratio of the posterior standard deviations of PAA versus  RW-$\lambda$
across all simulated datasets. We observe that for most interactions this ratio is distributed around one.
For the last four interactions, where the latent is involved, we observe that PAA has systematically  lower standard deviation.

\subsection{ Torus Mandibularis in Eskimoid Groups Data
\label{ex}}

We illustrate the proposed methodology by using the dataset of \cite{MullerMayhall}
studying the incidence of the morphological trait torus mandibularis in different Eskimo groups.
Torus mandibularis is a bony growth in
the mandible along the surface nearest to the tongue. This
morphological structure of the month is frequently used by
anthropologist to study differences among populations and among
groups within the
 same population.  This data have been previously analysed by \cite{Bishop_et_al} via
 log linear models, and by \cite{Lupparelli} via marginal log-linear graphical models.

 For our analysis, we consider the  data presented in Table \ref{torus}, cross-classifying  age (A),
  incidence of torus mandibularis (I), sex (S) and  population (P).
 The dataset is a dichotomized version of the original data of \cite{MullerMayhall}.
 The examined Eskimo groups refers to different geographical regions, Igloolik and Hall Beach groups are from Foxe Basin area of Canada whereas Aleut are from Western
Alaska. Furthermore, the data of the Aleuts group were collected
by an investigator different from the one who collected the data
for the first two groups, with a time difference between
investigations of about twenty years. For the previous reasons we
decided to reclassify the data in two groups: the first one including Igloolik and Hall Beach  and the second one Aleut.
Finally, variable age has been classified in two groups according
to the median value.

\begin{table}[hp!]\caption{Torus Mandibularis in Eskimo Populations} \label{torus}
\begin{center}
\begin{tabular}{lllll}
\hline
&  &  & \multicolumn{2}{c}{ {\it Age Groups} (A)} \\ \cline{4-5}
{\it Population} (P) & {\it Sex} (S) & {\it Incidence} (I) & 1-20 & Over 20 \\
\cline{1-5}
 Igloolik and Hall Beach &
\begin{tabular}{ll}
Male \\
\\
Female \\
\end{tabular}
&
\begin{tabular}{l}
Present \\
Absent \\
Present \\
Absent%
\end{tabular}
&
\begin{tabular}{l}
19 \\
103 \\
16 \\
87%
\end{tabular}
&
\begin{tabular}{l}
73 \\
38 \\
61 \\
36%
\end{tabular}
\\
Aleut &
\begin{tabular}{l}
Male \\
\\
Female \\
\end{tabular}
&
\begin{tabular}{l}
Present \\
Absent \\
Present \\
Absent%
\end{tabular}
&
\begin{tabular}{l}
6 \\
19 \\
4 \\
17%
\end{tabular}
&
\begin{tabular}{l}
18 \\
14 \\
10 \\
20
\end{tabular}
\\ \hline
\end{tabular}
\end{center}
\end{table}

\begin{figure}[h]
\caption{Bi-directed graph for Torus data}
\label{torus1}
  \centering  \includegraphics[width=3cm]{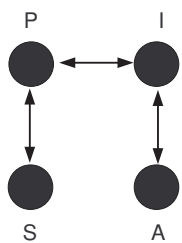}%
          \end{figure}

From the analysis of \cite{Lupparelli} we know that the model represented in
Figure \ref{torus1}
fits well the original  data, hence, in the following, we
concentrate on this specific four-chain graph.

Table \ref{MCMC} reports the posterior means and standard
deviations for the marginal log-linear interactions obtained via
$10000$ iteration with a burn in of $1000$ with the proposed
 PAA algorithm and the RW-$\lambda$.
The maximum likelihood estimates (MLEs) and corresponding approximate standard errors are also reported for comparative purposes.  MLEs are obtained implementing an algorithm for the optimization of the Lagrangian log-likelihood which includes a set of zero constraints satisfying the marginal independence model; see \cite{Lupparelli} for details. Standard errors for the parameter estimates can be simply derived from the estimate of the Hessian matrix of the Lagrangian log-likelihood; see \cite{AS}.  

\begin{table}[hp]\caption{Posterior summaries and MLEs for marginal log-linear interactions for the Torus Mandibularis data} \label{MCMC}
\begin{center}
\begin{tabular}{|l|c|c|c|c|c|c|}
  \hline
      & \multicolumn{2}{c|}{PAA} & \multicolumn{2}{c|}{RW-$\lambda$}   & \multicolumn{2}{c|}{ML}\\
   & Mean & SD & Mean & SD & Estimate & SE \\
  \hline
  $\lambda^{AP}_\emptyset$    &-1.391 & 0.004 &-1.391 & 0.004 & & \\
  $\lambda^{AP}_A(2)$         &-0.001 & 0.042 &-0.003 & 0.043 &-0.002 & 0.043\\
  $\lambda^{AP}_P(2)$         &-0.072 & 0.043 &-0.079 & 0.043 &-0.072 & 0.043\\
  $\lambda^{AP}_{AP}(2,2)$    & 0.000 & 0.000 & 0.000 & 0.000 & 0.000 & \\
  $\lambda^{AS}_S(2)$         &-0.697 & 0.053 &-0.695 & 0.055 &-0.699 & 0.054\\
  $\lambda^{AS}_{AS}(2,2)$    & 0.000 & 0.000 & 0.000 & 0.000 & 0.000 &\\
  $\lambda^{IS}_I(2)$         & 0.234 & 0.045 & 0.241 & 0.044 & 0.232 & 0.044 \\
  $\lambda^{IS}_{IS}(2,2)$    & 0.000 & 0.000 & 0.000 & 0.000 & 0.000 & \\
   $\lambda^{APS}_{P S}(2,2)$   & 0.004 & 0.053 &-0.009 & 0.055 & 0.003 & 0.054\\
  $\lambda^{APS}_{APS}(2,2,2)$ & 0.000 & 0.000 & 0.000 & 0.000 & 0.000 & \\
  $\lambda^{AIS}_{AI}(2,2)$    &-0.509 & 0.051 &-0.505 & 0.052 &-0.507 & 0.051\\
   $\lambda^{AIS}_{AIS}(2,2,2)$& 0.000 & 0.000 & 0.000 & 0.000 & 0.000 & \\
 $\lambda^{AIPS}_{IP}(2,2)$     & 0.057 & 0.058 & 0.082 & 0.063 & 0.052 & 0.062\\
 $\lambda^{AIPS}_{AIP}(2,2,2)$  & 0.132 & 0.068 & 0.049 & 0.065 & 0.151 & 0.062 \\
  $\lambda^{AIPS}_{ISP}(2,2,2)$ & 0.029 & 0.041 & 0.066 & 0.063 & 0.072 & 0.062\\
  $\lambda^{AIPS}_{AIPS}(2,2,2)$& 0.047 & 0.046 & 0.034 & 0.063 & 0.037 & 0.062\\
  \hline
\end{tabular}
\end{center}
\end{table}

From Table \ref{MCMC} 
we observe that the posterior estimates (for both MCMC methods) and
the MLEs coincide for all interactions and main effects obtained by marginals where no latent variable is involved;   also Figure \ref{ergodic_plots} in the Appendix.
This is not the case for
 $\lambda^{AIPS}_{IP}(2,2)$,
 $\lambda^{AIPS}_{IPS}(2,2,2)$,
 $\lambda^{AIPS}_{AIPS}(2,2,2)$,
where the latent variable is involved. 
 More specifically, the posterior standard deviations are lower by  $6.5\%$, $34\%$ and $26\%$, respectively.
This result is intuitively expected since PAA moves across the correct posterior distribution
defined on the space of parameterisation with compatible marginal probabilities. On the other hand,
 both the RW-$\lambda$ and the approximate MLEs standard errors are obtained without considering the restrictions imposed in
order to obtain  parameterisations leading to compatible marginal probabilities.

Finally, differences are also observed for interaction  $\lambda^{AIPS}_{AIP}(2,2,2)$
where RW-$\lambda$ provides posterior means far away from the corresponding MLEs with PAA being quite closely
and standard deviance slightly higher than both the corresponding values of RW-$\lambda$ and the MLEs standard errors.



In terms of computational time,
 PAA was found to be faster with elapsed CPU time lower by $46\%$ compared with the corresponding one for RW-$\lambda$
(32.6 versus 60.8 seconds for 1000 iterations).
Equi\-va\-lently, the reported user CPU time was $48\%$ lower for PAA than the one for RW-$\lambda$
(29.9 versus 57.2 seconds for 1000 iterations).
All runs were performed in a  Windows 10 PC Intel Core i7-5500U CPU 2.4GHz with 16GB memory;
all timings were obtained with the {\sf system.time} function in {\sf R}.

For 11,000 iterations, the effective sample size (ESS) of the methods are of similar size ranging from
$43\%$ in favour of RW-$\lambda$ (for S main effect) to $36\%$ in favour of our method
(for the four-way interaction).

\begin{figure}[hbtp]
      \caption{Boxplots of the relative efficiency of Prior-Adjustment Algorithm (PAA) compared to the random walk MCMC obtained by {\sf CODA} and {\sf STAN}
			         for the Torus Mandibularis data}
      \vspace{-7em}
      \label{Box_Reff_example}
      \begin{center}
      \includegraphics{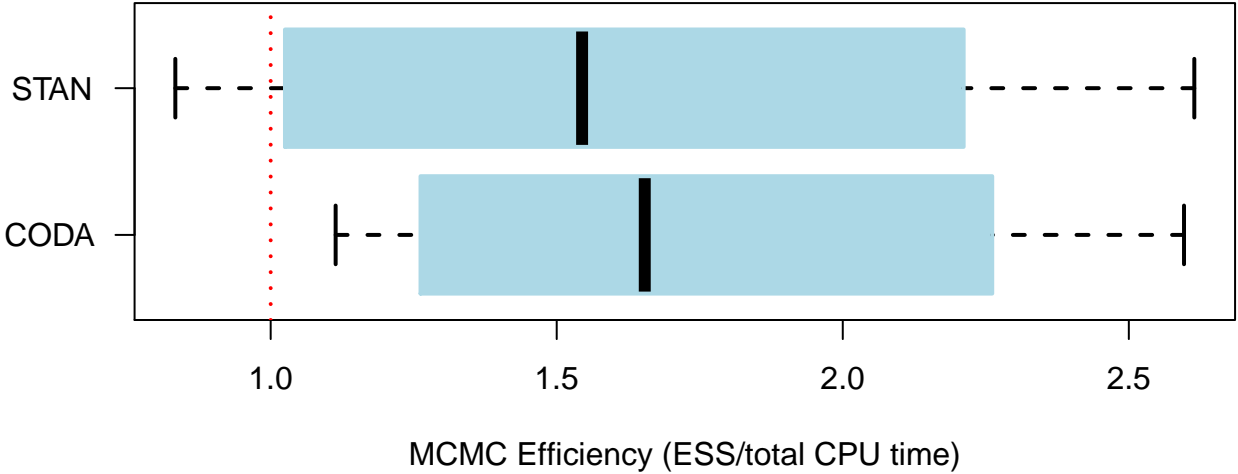}
        \end{center}
\end{figure}

Taking into consideration also the computational time of the two algorithms,
the MCMC efficiency (ESS/Elapsed CPU time) for the proposed algorithm is considerably higher for all interactions ranging from $2\%$ increase to $141\%$.
The average relative MCMC efficiency is higher for our method by $65\%$ compared with the one of the random walk MCMC.
The above statistics were calculated with the {\sf effectiveSize} function of {\sf coda} package in {\sf R}.
The overall picture is similar if we consider results using the  {\sf monitor} function of {\sf rstan} package in {\sf R}; see Figure \ref{Box_Reff_example} for a visual representation of the relative efficiency for all interactions.

Finally, the MCMC errors were lower for the majority of the interactions by using the naive estimator  of {\sf CODA} package with relative values ranging from $0.61$ up to  $1.13$
while the corresponding relative values for the time series based estimator are ranging from $0.66$ up to $1.30$.
The naive estimator of MCE is simply given by the usual standard error ignoring autocorrelation
(i.e. the MCMC estimated posterior standard deviation over the square root of the number of iterations).

\section{Discussion \label{disc}}

A possible way to parametrise discrete graphical models of marginal
independence is by using  the log-linear marginal models of \cite{BergsmaRudas}. The marginal log-linear interactions are calculated
from specific marginals of the original table, and independencies imply zero constraints on specific set of interactions in a similar manner as in conditional log-linear graphical models.

In this work we focus on the Bayesian estimation of the log-linear interactions for graphical models of marginal independence.
 In particular, the method we propose allows  us to assign prior directly on marginal log-linear interactions rather than on the probability interactions.
This facilitates the incorporation of prior information since several models of interest can be specified by zero/linear constraints on log-linear terms.
Bayesian analysis of such models is not widespread mainly due to the computational problems involved
in the derivation of their posterior distribution.
More specifically, MCMC methods need to be used since no conjugate analysis is available.
Major difficulties arise from the fact that we need to sample from a posterior distribution
defined on the space of parameterisations with compatible marginal probability distributions.
In the proposed   algorithms (PBIS and  PAA), we satisfy such restrictions by sampling from the probability space of the graphical model of marginal independence under consideration.
Then we transform the probability parameter values to the corresponding marginal log-linear ones
avoiding the iterative procedure needed for the evaluation of the likelihood.
In order to achieve this, we exploit the augmented DAG representation of the model.
This not only facilitates the prior elicitation but also the construction of the Jacobian matrix
involved in the acceptance probability of the induced Metropolis steps.
 PBIS is a more elaborate method but   it leads to an automatic setup for sampling log-linear interactions by the posterior distribution.
On the other hand, PAA can be implemented in a more straightforward manner. It leads to an efficient and fully automatic setup, which can be thought
as a approximate method for sampling from the posterior distribution. It further avoids any time-consuming and troublesome tuning of MCMC parameters.

For future research, the authors would like to exploit  and study the connections between the prior and the posterior distributions for the two different parameterisations (probability versus marginal log-linear).
Moreover, extension of the method to accommodate fully automatic selection, comparison and model averaging techniques is an intriguing topic for further investigation.

\section*{Supplementary Material}
Supplementary Material: Appendix for the Paper ``Probability Based Independence Sampler  for Bayesian Quantitative Learning in Graphical Log-Linear  Marginal Models'' (available at \url{http://www.stat-athens.aueb.gr/~jbn/papers/files/2016_Ntzoufras_Tarantola_Lupparelli_Supplementary.pdf}). The supplementary material includes details for the Jacobian calculations (Appendix \ref{app_jac}) 
and details for the construction of ${\mathbf M}$ and  ${\mathbf C}$ matrices
(Appendices \ref{calculation_M} and \ref{calculation_C} 
respectively).
Finally, some additional results for the illustrated examples of Section \ref{Examples} are
provided in Appendix \ref{add_results}.

\section*{Acknowledgments}
We would like to thank  Giovanni Marchetti for providing us the {\sf R} function  {\sf inv.mlogit}.
This research was partially funded by the Research Centre of the Athens University of Economics and
Business (Funding program for the research publications of the AUEB Faculty members) and by the
Department of Economics and Management of University of Pavia.


\newpage

\pagenumbering{roman}

\renewcommand\thefigure{\thesection.\arabic{figure}}
\setcounter{figure}{0}

\begin{center}
{\bf \Large Supplementary Material: Appendix for the Paper ``Probability Based Independence Sampler  for Bayesian Quantitative Learning in Graphical Log-Linear  Marginal Models''}
\end{center}

\vspace{1em}
\begin{center}
\begin{tabular}{c@{~~~~}c@{~~~}c@{}c}
\large Ioannis Ntzoufras, &
\large Claudia Tarantola & and &
\large Monia Lupparelli \\
\footnotesize
Athens University of Economics &
\footnotesize University of Pavia, Italy &&
\footnotesize  University of Bologna, Italy  \\[-0.5em]
\footnotesize   \& Business, Greece &&&
\end{tabular}
\end{center}

\appendix

\section{Jacobian calculations}
\label{app_jac}

Equation (\ref{jacelements}) can be rewritten as
$$
\frac{\partial \lambda_k}{\partial  \Pi_\jmath}
=
\sum \limits_{l=1}^{ \textsc{c}_{C} } Q_{k l \jmath } \mbox{~~with~~}
Q_{k l \jmath } = C_{k l} \frac{ \sum \limits_{\imath=1}^{|{\cal I}|} M_{l \imath} \Delta_{\imath \jmath} }
                               { \sum \limits_{\imath=1}^{|{\cal I}|} M_{l \imath} P_\imath               },
$$
where
$\dn{M}=( M_{l \imath} )$ is a $\textsc{c}_{C} \times |\cal{I}|$ matrix,
$\dn{\Delta}= ( \Delta_{\imath \jmath} )$ is a matrix of dimension $|{\cal I}|  \times d_\Pi$,
and $d_\Pi$ is the dimension of the parameter vector $\dn{\Pi}$.

\subsection{Step-by-step computation of the Jacobian matrix}


Once $\dn{\Delta}$ is obtained (see Appendix \ref{sec_delta_comp}), we can construct the Jacobian matrix using
the following steps

\begin{enumerate}

\item construct $\dn{H}=\dn{M} \dn{\Delta}$   a matrix of dimension $\textsc{c}_{C} \times  dim( \Pi)$;

\item construct $\dn{\Gamma}=\dn{M} \dn{P}$  vector of dimension $\textsc{c}_{C}\times 1$;

\item
construct $\dn{\Gamma}^{\prime}= \dn{\Gamma}\dn{1}_{d_\Pi}^T$, with
$\dn{1}_{d}$ a $d_\Pi \times 1$ vector of ones,
and  $\dn{\Gamma}^{\prime}$  a matrix of dimension $\textsc{c}_{C} \times d_\Pi$ with all columns equal to $\dn{\Gamma}$;


\item set $\dn{H}^{\prime}=\dn{H} \circ  \dn{\Gamma}^{\prime\prime}$
where $\circ$ indicates the Hadamard product (element by element
multiplication), and $\dn{\Gamma}^{\prime\prime}$ is a matrix with
elements $\Gamma^{\prime\prime}_{\nu \kappa} = 1/\Gamma^{\prime}_{\nu
\kappa}$;

\item denote with $\dn{J} = \dn{C} \dn{H}^{\prime}$  the Jacobian matrix dimension $d_\Pi \times d_\Pi$.

\end{enumerate}


\subsection{Computational details for the elements of $\dn{\Delta}$}
\label{sec_delta_comp}


We now describe how we can analytically obtain the elements of
matrix $\dn{\Delta}$ defined in (\ref{delta}). Each element of
this matrix is derivative of type $\dfrac{\partial p(i)}{\partial
\pi_{u | \pa(u)} \big( j_u | j_{\pa(u)} \big) }$, $i \in {\cal I}$ and $j \in {\cal I}^{\cal A}$.
We can separate
the computations in two separate cases:
\begin{description}
\item [Case A:]  $u \in {\cal V}$, i.e. $u$ is an observable variable.
 \item [Case B:]  $u \in {\cal L}$, i.e. $u$ is a latent variable.
\end{description}

\subsubsection{\textbf{Case A}: Computations for Expression (\ref{jac1a})}

For any variable $u \in {\cal V}$ and $j_u < |{\cal I}_u|$ we have that
\begin{eqnarray}
  \dfrac{\partial  p(i) } { \partial \pi_{u|\paua} \big(j_u| j_{\paua}\big) }
= \dfrac{\partial  \sum \limits_{ i_{\cal L} \in {\cal I}_{\cal L} } p^{\cal A} \big(i, i_{\cal L} \big) }
        {\partial \pi_{u|\paua} \big(j_u| j_{\paua}\big) }
= \sum_{ \iaug_{\cal L} \in {\cal I}_{{\cal L}}} \dfrac{\partial  p^{\cal A} \big( \iaug \big) }
                                             {\partial \pi _{u|\paua} \big(j_u| j_{\paua} \big) } \nonumber
\end{eqnarray}
since $\iaug = (i, i_{\cal L}) \Leftrightarrow \iaug_{\cal V}=i$
and $\iaug_{\cal L}=i_{\cal L}$.
 From (\ref{augm_joint_probs}) we
further obtain that
\begin{eqnarray}
\lefteqn{
\dfrac{\partial  p(i) } { \partial \pi_{u|\paua} \big(j_u| j_{\paua}\big) }  = } \nonumber \\
&=& \sum_{ \iaug_{\cal L} \in {\cal I}_{{\cal L}}}
                      \frac{\partial \left\{ \pi_{u|\paua} \big( \iaug_u| \iaug_{\paua} \big)
                  \prod \limits _{v \in {\cal V}\cup {\cal L}\setminus u}\  \pi_{v|\pava} \big(\iaug_v| \iaug_{\pava} \big)   \right\}}
            {\partial \pi_{u|\paua} \big(j_u| j_{\paua}\big) } =  \nonumber\\
 &=& \sum_{ \iaug_{\cal L} \in {\cal I}_{{\cal L}}}
  \left\{
   \left[ \prod \limits _{v \in {\cal V}\cup {\cal L}\setminus u}\
           \pi_{v|\pava} \big(\iaug_v| \iaug_{\pava } \big) \right]
   \times
   \frac{\partial  \pi_{u|\paua} \big(\iaug_u|\iaug_{\paua } \big)    }{\partial \pi_{j_u|\paua} \big(j_u|j_{\paua } \big)}
 \right\}=
\nonumber\\
&=& \sum_{ \iaug_{\cal L} \in {\cal I}_{{\cal L}}} \left\{
    \left[ \prod \limits _{v \in {\cal V}\cup {\cal L}\setminus u}\
           \pi_{v|\pava} \big(\iaug_v| \iaug_{\pava \setminus {\cal L}}, \iaug_{{\cal L}_v} \big) \right]
   \times \frac{\partial  \pi_{u|\paua} \big(\iaug_u|\iaug_{\paua \setminus {\cal L}}, \iaug_{{\cal L}_u} \big)    }
            {\partial \pi_{u|\paua} \big(j_u| j_{\paua \setminus {\cal L}}, j_{{\cal L}_u}\big) } \right\},
                        \label{eq_for_derivative}
\end{eqnarray}
where ${\cal L}_u={\cal L} \cap \paua$, is the set latent variables ${\cal L}$
 that are parents of $u$.
In the following we indicate with
by $\overline{{\cal L}_u} = {\cal L} \setminus \paua$ the  latent variables that are not parents of $u$.

Therefore, (\ref{eq_for_derivative}) can be rewritten
\begin{eqnarray*}
\lefteqn{ \dfrac{\partial  p(i) }
     {\partial \pi_{u|\paua} (j_u| j_{\paua}) } =}\\
&=& \sum_{ \iaug_{\overline{{\cal L}_u}} \in {\cal I}_{\overline{{\cal L}_u}}}
\left\{
    \sum_{ \iaug_{{\cal L}_u} \in {\cal I}_{{\cal L}_u} }  \left[
    \prod \limits _{v \in {\cal V}\cup ({\cal L}_u \cup \overline{{\cal L}_u}) \setminus u}\
    \pi_{v|\pava} \big(\iaug_v| \iaug_{\pava \setminus {\cal L}}, \iaug_{{\cal L}_v} \big)
\times \frac{\partial \pi_{u|\paua} \big(\iaug_u| \iaug_{\paua \setminus {\cal L}}, \iaug_{{\cal L}_u} \big)
}
     {\partial \pi_{u|\paua} \big(j_u| j_{\paua \setminus {\cal L}}, j_{{\cal L}_u} \big)  }  \right] \right\} \\
&=&\sum_{ \iaug_{\overline{{\cal L}_u}} \in {\cal I}_{\overline{{\cal L}_u}}}
\left\{
    \sum_{ \iaug_{{\cal L}_u} \in {\cal I}_{{\cal L}_u} }  \left[
                        \prod \limits _{v \in  \overline{{\cal L}_u}}\ \pi_{v|\pava} \big(\iaug_v| \iaug_{\pava \setminus {\cal L}}, \iaug_{{\cal L}_v} \big) \right. \right.\\
       && \hspace{2.5cm} \left. \left.
       \times \prod \limits _{v \in {\cal V}\cup {\cal L}_u\setminus u}\ \pi_{v|\pava} \big(\iaug_v| \iaug_{\pava \setminus {\cal L}}, \iaug_{{\cal L}_v} \big)
       \times \frac{\partial \pi_{u|\paua} \big(\iaug_u| \iaug_{\paua \setminus {\cal L}}, \iaug_{{\cal L}_u} \big)}
                   {\partial \pi_{u|\paua} \big(j_u| j_{\paua \setminus {\cal L}}, j_{{\cal L}_u} \big)  }  \right] \right\} \\
&=&  \sum_{ \iaug_{\overline{{\cal L}_u}} \in {\cal I}_{\overline{{\cal L}_u}}}
\left\{
  \left[  \prod \limits _{v \in  \overline{{\cal L}_u}}\
    \pi_{v|\pava} \big(\iaug_v| \iaug_{\pava \setminus {\cal L}}, \iaug_{{\cal L}_v} \big)  \right]\right.  \\
    &&\left. \times
    \sum_{ \iaug_{{\cal L}_u} \in {\cal I}_{{\cal L}_u} }\left(
    \prod \limits _{v \in {\cal V}\cup {\cal L}_u\setminus u}\
    \pi_{v|\pava} \big(\iaug_v| \iaug_{\pava \setminus {\cal L}}, \iaug_{{\cal L}_v} \big)
 \times \frac{\partial \pi_{u|\paua} \big(\iaug_u| i_{\paua \setminus {\cal L}}, \iaug_{{\cal L}_u}\big)  }
     {\partial \pi_{u|\paua} \big(j_u| j_{\paua \setminus {\cal L}}, j_{{\cal L}_u} \big)  }       \right)  \right\}.
 (\arabic{equation}) \label{der2}\addtocounter{equation}{1}
\end{eqnarray*}

We now concentrate on the second line of (\ref{der2}), that is
\begin{eqnarray*}
\lefteqn{
    \sum_{ \iaug_{{\cal L}_u} \in {\cal I}_{{\cal L}_u} }\left( \prod \limits _{v \in {\cal V}\cup {\cal L}_u\setminus u}\
    \pi_{v|\pava} \big(\iaug_v| \iaug_{\pava \setminus {\cal L}}, \iaug_{{\cal L}_v} \big)
    \times
    \frac{\partial \pi_{u|\paua} \big(\iaug_u| \iaug_{{\cal L}_u}, \iaug_{\paua \setminus {\cal L}} \big)  }
         {\partial \pi_{u|\paua} \big(j_u| j_{{\cal L}_u}, j_{\paua \setminus {\cal L}} \big)  }       \right) =} \\
&&\hspace{1cm} = \begin{cases}
~~~\prod \limits _{v \in {\cal V}\cup {\cal L}_u\setminus u}\
\pi_{v|\pava} \big(\iaug_v| \iaug_{\pava} \big)
& \mbox{~if~} \iaug_u = j_u<|{\cal I}_u| \mbox{~and~ } \iaug_{\paua}= j_{\paua}  \\
-\prod \limits _{v \in {\cal V}\cup {\cal L}_u\setminus u}\
\pi_{v|\pava} \big(\iaug_v| \iaug_{\pava} \big)
& \mbox{~if~} j_u \neq \iaug_u = |{\cal I}_u| \mbox{~and~ } \iaug_{\paua}= j_{\paua}  \\
~~~~~~~~0 & \mbox{~if~} j_u \neq \iaug_u<|{\cal I}_u| \mbox{~or~ } \iaug_{\paua} \neq j_{\paua}  \\
\end{cases} \, .     \\
\end{eqnarray*}

In the case where $\iaug_u = j_u<|{\cal I}_u|$ and $i_{\paua\setminus
{\cal L}} = j_{\paua\setminus {\cal L}}$ then
\begin{eqnarray*}
\lefteqn{
\dfrac{\partial  p(i) }
     {\partial \pi_{u|\paua} \big(j_u| j_{\paua}\big) } =}\\
&=&  \sum_{ \iaug_{\overline{{\cal L}_u}} \in {\cal I}_{\overline{{\cal L}_u}}}
\left\{
    \left[ \prod \limits _{v \in  \overline{{\cal L}_u}}\   \pi_{v|\pava} \big(\iaug_v| \iaug_{\pava} \big) \right]
    \prod \limits _{v \in {\cal V}\cup {\cal L}_u\setminus u}\ \pi_{v|\pava} \big(\iaug_v| \iaug_{\pava} \big)
    I_{ \{\iaug_{\paua}= j_{\paua} \} }
     \right\}     \\
&=&  \sum_{ \iaug_{\overline{{\cal L}_u}} \in {\cal I}_{\overline{{\cal L}_u}}} \left\{
    \prod \limits _{v \in   {\cal V}\cup {\cal L} \setminus u}\ \pi_{v|\pava} \big(\iaug_v| \iaug_{\pava}  \big)
     I_{ \{\iaug_{\paua}= j_{\paua} \}} \right\}     \\
&=&  \sum_{ \iaug_{\overline{{\cal L}_u}} \in {\cal I}_{\overline{{\cal L}_u}}}
\left\{
    \dfrac{ \pi_{u|\paua} ( \iaug_u | \iaug_{\paua} ) }
          { \pi_{u|\paua} ( \iaug_u | \iaug_{\paua} ) }
    \prod \limits _{v \in   {\cal V}\cup {\cal L} \setminus u}\ \pi_{v|\pava} \big(\iaug_v| \iaug_{\pava} \big)
    I_{ \{\iaug_{\paua}= j_{\paua} \}}
 \right\}   \\
&=&  \sum_{ \iaug_{\overline{{\cal L}_u}} \in {\cal I}_{\overline{{\cal L}_u}}}
\left\{
    \dfrac{ p^{\cal A} ( i, j_{{\cal L}_u}, j_{\overline{{\cal L}_u}}  ) }
         { \pi_{u|\paua} ( \iaug_u | j_{\paua} ) } \right\}
=
    \frac{ \sum \limits_{ \iaug_{\overline{{\cal L}_u}} \in {\cal I}_{\overline{{\cal L}_u}}}
                        p^{\cal A} ( i, j_{{\cal L}_u}, j_{\overline{{\cal L}_u}}  ) }
         { \pi_{u|\paua} ( \iaug_u | j_{\paua} ) } \, .
\end{eqnarray*}

Finally we obtain that
\be
\dfrac{\partial  p(i) }
     {\partial \pi_{u|\paua} \big(j_u| j_{\paua}\big) }
=\frac{  p^{{\cal A}_u} ( i,  j_{{\cal L}_u}   ) }
         { \pi_{u|\paua} ( \iaug_u | j_{\paua} ) } \, .
\label{case_a_res1}
\ee

In a similar way, we find that
\be
\dfrac{\partial  p(i) }
     {\partial \pi_{u|\paua} (j_u| j_{\paua}) } =
     -  \frac{  p^{{\cal A}_u} ( i,  j_{{\cal L}_u}   ) }
         { \pi_{u|\paua} ( \iaug_u | j_{\paua} ) }
\label{case_a_res2}
\ee
for $j_u \neq \iaug_u = |{\cal I}_u| $
and $i_{\paua\setminus {\cal L}} = j_{\paua\setminus {\cal L}}$.

Finally, if $j_u \neq \iaug_u<|{\cal I}_u|\mbox{~or~ } i_{\paua\setminus {\cal L}}
\neq j_{\paua\setminus {\cal L}}$, (\ref{der2}) will become equal to
\be
\dfrac{\partial  p(i) } {\partial \pi_{u|\paua} (\iaug_u| j_{\paua}) } = 0.
\label{case_a_res3}
\ee

Since $u$ is an observed variable $i_u\equiv \iaug_u$, from equations (\ref{case_a_res1})--(\ref{case_a_res3}), we obtain the
final expressions 

$$
\dfrac{\partial  p(i) }
     {\partial \pi_{u|\paua} \big(j_u| j_{\paua} \big) } =
     \delta( i, j)
     \frac{  p^{ {\cal A}_u } \big( i, j_{ {\cal L}_u }   ) } { \pi_{u|\paua} ( i_u  |  j_{\paua} \big) }
$$
with
$$
\delta( i, j) =
\begin{cases}
~~~1  & \mbox{~if~} i_u = j_u<|{\cal I}_u|$ and $i_{\paua\setminus {\cal L}} = j_{\paua\setminus {\cal L}}\vspace{0.5em} \\
  -1  & \mbox{~if~} j_u \neq i_u = |{\cal I}_u| $ and $i_{\paua\setminus {\cal L}} = j_{\paua\setminus {\cal L}}\\
~~~0  &  \mbox{~if~} j_u \neq i_u < |{\cal I}_u| $ or
$\;\;i_{\paua\setminus L} \neq j_{\paua\setminus {\cal L}}
\end{cases}~,
$$
where
$$
p^{ {\cal A}_u } \big( i, j_{ {\cal L}_u} \big) =
\begin{cases}
P \Big( Y_{\cal V} = i,  Y_{{\cal L} \cap \pa(u)} = j_{{\cal L} \cap \pa(u)} \Big) & {\cal L}_u \neq \emptyset \\
p(i) & {\cal L}_u =\emptyset
\end{cases}.
$$

\subsubsection{\textbf{Case B}: Computations for Expression (\ref{jac2})}

For every latent variable $u \in {\cal L}$ the parent set is empty.
Hence
$\pi_{u|\paua} \big( \iaug_u| \iaug_{\paua} \big) = \pi_{u} \big( \iaug_u\big)$
for all $\iaug \in {\cal A}$.
Furthermore, it holds that
 $p(i)=\sum \limits_{ i_{\cal L} \in {\cal I}_{\cal L}} p^{\cal A}(i, i_{\cal L})=\sum \limits_{ \iaug_{\cal L} \in {\cal I}_{\cal L}} p^{\cal A}(\iaug)$~.

Therefore, for any parameter $\pi_{u}(j_{u})$ with $j_{u} \neq |{\cal I}_u|$ and $u \in {\cal L}$,
the derivative is given by
\begin{eqnarray*}
\frac{\partial p(i) }{\partial \pi_{u} (j_u ) }
&= & \frac{\partial  \sum \limits_{ i_{\cal L} \in {\cal I}_{\cal L}} p^{\cal A}(i, i_{\cal L}) }
          {\partial \pi_{u  } (j_u) }=
          \frac{\partial  \sum \limits_{ \iaug_{\cal L} \in {\cal I}_{\cal L}} p^{\cal A}(\iaug) }
          {\partial \pi_{u  } (j_u) }
 =  \sum \limits_{ \iaug_{\cal L} \in {\cal I}_{\cal L}} \frac{\partial   p^{\cal A}( \iaug ) }
          {\partial \pi_{u  } (j_u) } \\
&=&         \sum_{ \iaug_{{\cal L}\setminus u} \in {\cal I}_{{\cal L}\setminus u}}
    \left\{ \sum_{ \iaug_u \in {\cal I}_{u}}
    \dfrac{\partial  \left\{ \pi_{u  } (\iaug_u )
                            \prod \limits _{v \in {\cal V}\cup {\cal L}\setminus \{u\}}\
                                        \pi_{v|\pava} \left(\iaug_v| \iaug_{\pava} \right)   \right\}  }
          {\partial \pi_{u  } (j_u) }     \right\} \\
&=&  \sum_{ \iaug_{{\cal L}\setminus u} \in {\cal I}_{{\cal L}\setminus u}}
    \left\{ \sum_{ \iaug_u \in {\cal I}_{u}} \left[
    \dfrac{\partial   \pi_{u  } (\iaug_u ) } {\partial \pi_{u  } (j_u) }
    \times \prod \limits _{v \in {\cal V}\cup {\cal L}\setminus \{u\}}\
                                        \pi_{v|\pava} \left(\iaug_v| \iaug_{\pava} \right) \right] \right\}.
\end{eqnarray*}
The derivatives involved in the above equation are given by
\begin{eqnarray*}
\dfrac{\partial   \pi_{u  } (\iaug_u ) } {\partial \pi_{u  } (j_u) } =
\begin{cases}
~~~1 & \mbox{if }j_u=\iaug_u <|{\cal I}_u| \\
  -1 & \mbox{if }j_u\neq\iaug_u =|{\cal I}_u| \\
~~~0 & \mbox{if }j_u\neq\iaug_u <|{\cal I}_u|
\end{cases}     ~.
\end{eqnarray*}
Hence, we obtain
\begin{eqnarray*}
\frac{\partial p(i)}{\partial \pi_{u} (j_u ) }
&=& \sum_{ \iaug_{{\cal L}\setminus u} \in {\cal I}_{{\cal L}\setminus u}}
    \bigg\{
    \prod \limits _{v \in {\cal V}\cup {\cal L}\setminus \{u\}}\
                                        \pi_{v|\pava} (\iaug_v| \iaug_{\pava} ) I_{\{\iaug_u=j_{u}\}} \\
        &&
        \hspace{3.7em} -\prod \limits _{v \in {\cal V}\cup {\cal L}\setminus \{u\}}\
        \pi_{v|\pava} (\iaug_v| \iaug_{\pava} ) I_{ \{ \iaug_{u} = |{\cal I}_{u}| \} } \bigg\}   \\
&=& \sum_{ \iaug_{{\cal L}\setminus u} \in {\cal I}_{{\cal L}\setminus u}}
    \bigg\{
    \dfrac{\pi_u(\iaug_u)}{\pi_u(\iaug_u)}
    \prod \limits _{v \in {\cal V}\cup {\cal L}\setminus \{u\}}\
                                        \pi_{v|\pava} (\iaug_v| \iaug_{\pava}) I_{\{\iaug_u=j_{u}\}} \\
        &&
        \hspace{3.7em}
        -\dfrac{\pi_u(\iaug_u )}{\pi_u(\iaug_u )}
        \prod \limits _{v \in {\cal V}\cup {\cal L}\setminus \{u\}}\
        \pi_{v|\pava} (\iaug_v| \iaug_{\pava} )I_{ \{ \iaug_{u} = |{\cal I}_{u}| \} } \bigg\}   \\
&=&\sum_{ \iaug_{{\cal L}\setminus u} \in {\cal I}_{{\cal L}\setminus u}}
    \bigg\{
      \dfrac{ p^{\cal A}( \iaug_{\cal V}, j_u, \iaug_{{\cal L} \setminus u})}{\pi_u(j_u)}
     -\dfrac{ p^{\cal A}( \iaug_{\cal V}, |{\cal I}_u|, \iaug_{{\cal L} \setminus u}) }{ \pi_u( |{\cal I}_u| ) } \bigg\}   \\
&=&  \dfrac{ p^{{\cal A}_u}( \iaug_{\cal V}, j_u ) }{ \pi_u( j_u)}
    -\dfrac{ p^{{\cal A}_u}( \iaug_{\cal V}, |{\cal I}_u|) }{ \pi_u( |{\cal I}_u|) }\\
    &=&  \dfrac{ p^{{\cal A}_u}( i, j_u ) }{ \pi_u( j_u)}
    -\dfrac{ p^{{\cal A}_u}( i, |{\cal I}_u|) }{ \pi_u( |{\cal I}_u|) },
\end{eqnarray*}
where  $i=\iaug_{\cal V}$ and this concludes the computation of (\ref{jac2}).

\section{Construction of Matrix $\mathbf M$ \label{calculation_M}}

Let ${\cal M}=\Big\{ M_1, M_2, \dots, M_{|{\cal M}|} \Big\}$ be the set of marginals under consideration
by a given marginal log-linear parameterisation.
Let  ${\mathbf B}$ be a  binary matrix of
dimension $|\cal M| \times |{\cal V}|$ with elements $B_{m v}$
indicating whether a variable $v$ belongs to a specific marginal
$M_m$. The rows of ${\mathbf B}$ correspond to the
 marginals in ${\cal M}$ whereas the columns to the variables. The variables follow a reverse
 ordering, that is  column 1  corresponds to variable $Y_{|\cal V|}$, column 2 to variable
  $Y_{|{\cal V}|-1}$ and so on.
Matrix $\mathbf   B$ has elements
$$
B_{m v}=\left\{
\begin{array}{rl}
   1 &  \mbox{ if   $v \in M_m $ } \\
   0 &  \mbox{ otherwise.}
\end{array}
\right.
$$
for every $v \in \cal V$.

The marginalisation matrix $\mathbf M$ can be obtained applying
the following rules.
\begin{enumerate}
\item
     For each marginal $M_m $ and  each variable  $v$  we construct the following matrix
$$
 {\mathbf   A}_{m,v}=\left\{
\begin{array}{rl}
   {\mathbf   I}_{ |{\cal I}_v| } & \mbox{ if  }\; \;  B_{m v} = 1\\
   {\mathbf   1}_{ |{\cal I}_v|  }^T &\mbox{ if  } \; \;B_{m v} = 0
\end{array},
\right.
$$
        where
        ${\mathbf   I}_{|{\cal I}_v|}$ is the identity matrix of dimension $|{\cal I}_v|\times |{\cal I}_v|$ and
        ${\mathbf   1}_{|{\cal I}_v|}$ is a vector of dimension $|{\cal I}_v|\times 1$ with all elements equal to one.

 The probability vector of the
marginal table corresponding to  $M_m $ is given by $\mathbf M_m \mathbf
\pi$;
            where $\mathbf M_m$ is calculated as a Kronecker product of matrices ${\mathbf  A}_{m v}$
        \begin{eqnarray*}
        \mathbf   M_m
                &=& \bigotimes _{q=0}^{|{\cal V}|-1} {\mathbf   A}_{m,|{\cal V}|-q} \\
                &=&  {\mathbf   A}_{m,|{\cal V}|} \bigotimes {\mathbf   A}_{m,|{\cal V}|-1}
                                                    \bigotimes\dots \bigotimes{\mathbf   A}_{m,2} \bigotimes {\mathbf   A}_{m,1},
        \end{eqnarray*}
                where $\bigotimes$ denotes the Kronecker product which is implemented in reverse lexicographical order
                (starting from the last on in ${\cal V}$).

\item Matrix $\mathbf   M$ is constructed by stacking all the
$\mathbf M_m$ matrices
                $$
        \mathbf M = \left(  \begin{array}{l} \mathbf M_1 \\ \vdots \\ \mathbf M_m  \\ \vdots \\ \mathbf M_{|{\cal M}|} \end{array}  \right)~.
        $$
\end{enumerate}

%

\section{Construction of  Matrix $\mathbf C$ \label{calculation_C}}

Firstly we need to construct the design matrix ${\mathbf X}_{\cal V}$ for the saturated model for the cross-classification of discrete variables $Y_{\cal V}$ with sum to zero constraints.
Firstly, for each variable  $v$  we construct the following matrix
$${\mathbf   J}_{v}=\left(
\begin{array}{ll}
        1             & - \mathbf{1}^T_{(|{\cal I}_v|-1)} \\
\mathbf{1}_{(|{\cal I}_v|-1)} & ~~~\mathbf{I}_{(|{\cal I}_v|-1)}
\end{array}\right) ~,
$$
where ${\mathbf 1}_\kappa$ is a vector of ones of length $\kappa$
while $\mathbf I_{\kappa}$ is the identity matrix of dimension $\kappa \times \kappa$.

The design matrix of the satured model will be of dimension
$\left( \prod \limits _{v \in {\cal V}} | {\cal I}_v | \right) \times
 \left( \prod \limits _{v \in {\cal V}} | {\cal I}_v | \right)$
and  can be obtained as
$$
\mathbf Y_{\cal V} =\bigotimes _{q=0}^{|{\cal V}|-1} {\mathbf   J}_{|{\cal V}|-q} ~.
$$

The contrast matrix  $\mathbf C$ can be constructed by  using the
following rules.

\begin{enumerate}
\item For each margin $M_m $, we  construct the design matrix
${\mathbf X}_{M_m}$ corresponding to the saturated model (using  sum-to-zero constraints)
of the cross-classification of variables $Y_{M_m}$.
Then we consider its inverse  ${\mathbf X}_{M_m}^{-1}$ in order to obtain the corresponding contrast matrix.
Now, let $\mathbf C_m$ be a sub-matrix of the contrast matrix ${\mathbf X}_{M_m}^{-1}$ obtained by deleting rows
 corresponding to   interactions that are not obtained from margin $M_m$.

\item The contrast matrix $\mathbf C$ is obtained as
$$
\mathbf C = \bigoplus_{m:~ M_m \in {\cal M}}  \mathbf C_m=
\mbox{diag}(C_1, \ldots, C_{|\cal M|})~,
$$ where $\bigoplus$  denotes the  matrix direct  sum.

\end{enumerate}

\clearpage
\newpage
\section{Additional Results}
\label{add_results}

\begin{figure}[h!]
      \caption{Ergodic plots for marginal log-linear interactions using Prior-Adjustment Algorithm (PAA) compared
                                with approximate maximum likelihood estimates for the Torus Mandibularis data}\label{ergodic}
            \vspace{-2em}

      \label{ergodic_plots}
      \begin{center}

      \psfrag{A2}[c][b][0.7]{$\lambda_{\scriptscriptstyle A}^{\scriptscriptstyle AP}$}
      \psfrag{C2}[c][b][0.7]{$\lambda_{\scriptscriptstyle P}^{\scriptscriptstyle AP}$}
      \psfrag{B2}[c][b][0.7]{$\lambda_{\scriptscriptstyle I}^{\scriptscriptstyle IS}$}
      \psfrag{D2}[c][b][0.7]{$\lambda_{\scriptscriptstyle S}^{\scriptscriptstyle AS}$}

      \psfrag{C2D2}[c][b][0.7]{$\lambda_{\scriptscriptstyle P S}^{\scriptscriptstyle APS}$}
      \psfrag{A2B2}[c][b][0.7]{$\lambda_{\scriptscriptstyle A I}^{\scriptscriptstyle AIS}$}
      \psfrag{B2C2}[c][b][0.7]{$\lambda_{\scriptscriptstyle I S}^{\scriptscriptstyle IS}$~}
      \psfrag{A2D2}[c][b][0.7]{$\lambda_{\scriptscriptstyle A S}^{\scriptscriptstyle AS}$}

      \psfrag{A2B2C2}[c][b][0.7]{$\lambda_{\scriptscriptstyle A I P}^{\scriptscriptstyle AIPS}$}
      \psfrag{A2B2D2}[c][b][0.7]{$\lambda_{\scriptscriptstyle A I S}^{\scriptscriptstyle AIS}$}
      \psfrag{A2C2D2}[c][b][0.7]{$\lambda_{\scriptscriptstyle A P S}^{\scriptscriptstyle APS}$}
      \psfrag{B2C2D2}[c][b][0.7]{~~$\lambda_{\scriptscriptstyle I P S}^{\scriptscriptstyle AIPS}$}

      \psfrag{B2D2}[c][b][0.7]{~~$\lambda_{\scriptscriptstyle I S}^{\scriptscriptstyle IS}$}

      \psfrag{A2B2C2D2}[c][b][0.7]{~~~~~~~~$\lambda_{\scriptscriptstyle AISP}^{\scriptscriptstyle AISP}$}

      \includegraphics[width=13cm]{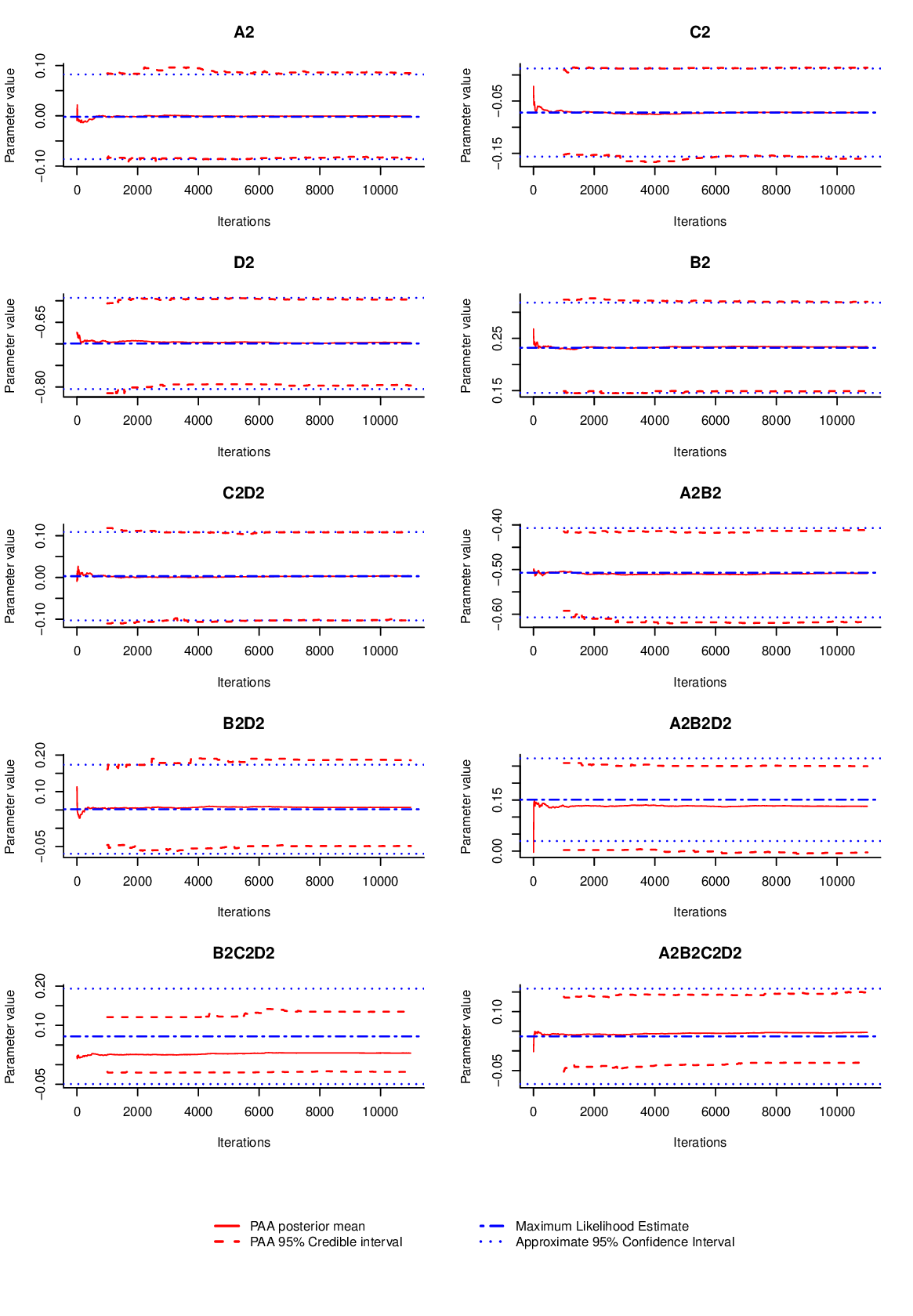}

        \end{center}
\end{figure}

Figure \ref{ergodic} depicts the posterior ergodic plots for the estimates  of the elements of $\vec{\dn{\lambda}}$ obtained using PAA and the corresponding approximate MLE based statistics
for the real dataset of Section \ref{ex}.

\end{document}